%% file: ocvirk.tex
\documentclass[useAMS,usenatbib,preprint2]{aastex}
\usepackage{amsmath,amssymb}
\usepackage{delarray}
\usepackage{graphicx}
\usepackage{url}
\usepackage{multirow}
\usepackage{rotating}
\usepackage{lscape}

\newcommand{\fesc}{\rm f_{esc}}
\newcommand{\xion}{\rm x_{ion}}
\newcommand{\zreion}{\rm z_{r}}
\newcommand{\zr}{\zreion}
\newcommand{\xsupp}{\rm x_{supp}}

\newcommand{\treion}{\rm t_{r}}

\newcommand{\Rvir}{\rm {R_{vir}}}
\newcommand{\mdm}{m_{\rm dm}}
\newcommand{\hmpc}{$\rm{h^{-1}Mpc}$ }
\newcommand{\hmkpc}{$\rm{h^{-1}kpc}$ }

\newcommand{\xm}{\langle \rm{x} \rangle^{\rm{m}}}

\newcommand{\mathd}{\mathrm{d}}

\newcommand{\Msun}{\ensuremath{\mathrm{M}_{\odot}}}
\newcommand{\hmo}{{\rm h^{-1}}}

\newcommand{\nicefrac}[2]{\leavevmode\kern.1em
            \raise.5ex\hbox{\the\scriptfont0 #1}\kern-.1em
      /\kern-.15em\lower.25ex\hbox{\the\scriptfont0 #2}}

\renewcommand{\arraystretch}{1.3}

\begin{document}

\title{The reionization of galactic satellite populations}
\shorttitle{ATON meets CLUES}

\author{P. Ocvirk\altaffilmark{1}, N. Gillet\altaffilmark{1}, D. Aubert\altaffilmark{1}, A. Knebe\altaffilmark{2}, N. Libeskind\altaffilmark{3}, J. Chardin\altaffilmark{1}, S. Gottl\"{o}ber\altaffilmark{3}, G. Yepes\altaffilmark{2}, Y. Hoffman\altaffilmark{4}}
\affil{
$^1$ Observatoire Astronomique de Strasbourg, Universit\'e de Strasbourg, CNRS UMR 7550, 11 rue de l'Universit\'e, 67000 Strasbourg, France}
\affil{
$^2$ Grupo de Astrof\'{i}sica, Departamento de Fisica Teorica, Modulo C-8, Universidad Aut\'{o}noma de Madrid, Cantoblanco E-280049, Spain}
\affil{
$^3$ Leibniz-Institute f\"{u}r Astrophysik Potsdam (AIP), An der Sternwarte 16, D-14482 Potsdam, Germany}
\affil{
$^4$ Racah Institute of Physics, Hebrew University, Jerusalem 91904, Israel}





\begin{abstract}
We use high resolution simulations of the formation of the local group post-processed by a radiative transfer code for UV photons, to investigate the reionization of the satellite populations of an isolated Milky Way-M31 galaxy pair in a variety of scenarios. We use an improved version of ATON which includes a simple recipe for radiative feedback.
In our baseline models, reionization is initiated by low mass, radiatively regulated haloes at high redshift, until more massive haloes appear, which then dominate and complete the reionization process. We investigate the relation between reionization history and {\em present-day} positions of the satellite population. We find that the average reionization redshift ($\zreion$) of satellites is higher near galaxy centers (MW and M31). This is due to the inside-out reionization patterns imprinted by massive haloes within the progenitor during the EoR, which end up forming the center of the galaxy. Thanks to incomplete dynamical mixing during galaxy assembly, these early patterns survive down to present day, resulting in a a clear radial gradient in the average satellites reionization redshift, up to the virial radius of MW and M31 and beyond. In the lowest emissivity scenario, the outer satellites are reionized about 180 Myr later than the inner satellites. 
This delay decreases with increasing source model emissivity, or in the case of external reionization by Virgo or M31, because reionization happens faster overall, and becomes spatially quasi-uniform at the highest emissivity.

\end{abstract}


%




\section{Introduction}

In the last decade, the epoch of reionization (hereinafter EoR) has received increasing attention. Most observational works now seem to converge on reionization beginning as early as z=15 \citep{kogut2003short} and finishing around z=6 \citep{fan2006}, in apparent agreement with theoretical predictions \citep{haardt2011}. The EoR also affects the way galaxies form: it has been suggested that the rising metagalactic UV radiation field is responsible for photo-evaporating the gas of low-mass galaxies \citep{gnedin2000,hoeft2006,benitez2014}, shutting down or delaying their star formation in the first billion years of the Universe.
This process could provide a credible solution to the ``missing satellites problem'' \citep{klypin1999,moore1999}, by inhibiting star formation in low mass galaxies at early times \citep{bullock2000,benson2002a,benson2002b,benson2003}.
In this framework, a number of simple semi-analytical models (hereinafter SAMs) have been shown to reproduce well the satellite population of the Milky Way (hereinafter MW), such as \cite{koposov2009,munoz2009,busha2010,maccio2010,li2010,font2011}. 
They suggest that the  ultra-faint dwarf galaxies (hereinafter UFDs) discovered by the SDSS \citep{martin2004,willman2005short,zucker2006,belokurov2007short, irwin2007short, walsh2007} are effectively reionisation fossils, living in sub-haloes of about $10^{6-9} \Msun$.
More recently, \cite{ocvirk2011,ocvirk2012a} showed that the structure of the UV background during reionization has a strong impact on the properties of the satellite population of galaxies. In particular, they showed that an internally-driven reionization led to significant changes in the radial distribution of satellites. 
It is therefore of prime importance to determine how central, or on the contrary how uniform the UV field is within a MW progenitor during reionization in a realistic setting. 
This is what the authors set out to determine in \cite{ocvirk2013} (hereinafter paper I). Thanks to the adequate spatial resolution of their radiative transfer (hereinafter RT) simulation (${\Delta x} = 21 \, {\rm{ h^{-1} kpc}}$), they were able to investigate the development and propagation of the ionization fronts (hereinafter I-front) within a typical Milky Way galaxy progenitor. They found that the process is patchy and dominated by a few (1-4) major regions expanding, percolating and finally filling the whole volume of the progenitor. The amount of structure in the process depends on the ionizing sources emissivity, and the patchiest reionization is obtained for the lowest emissivity.
Within this picture it becomes clear that the reionization redshift of a satellite depends on its position within the galaxy progenitor\footnote{Here the progenitor is defined as the volume containing all the particles which end up within 300 kpc of the MW center at z=0, as in \cite{ocvirk2013}. Therefore any MW satellite progenitor is already contained in the MW progenitor at high z.}. In particular, a future satellite which is close to the most massive halo of the MW progenitor is likely to reionize earlier than its more distant counterparts. Therefore it is reasonable to expect that the properties of the satellites will be correlated with their position at reionization. In this paper we ask: do these correlations survive down to present times? Should we expect the satellites properties to be correlated in some way with their z=0 distance to the MW's center for instance? From reionization to present times, the MW progenitor undergoes 12.7 Gyr of dynamical evolution, sometimes violent, thereby blurring the dynamical memory of the system. This process is expected to also blur or smoothen any correlation between satellite properties and their position. However, evolving gravitational collisionless systems are known to retain {\em some form} of memory of their past configuration due to Liouville's theorem \citep{gady}. This kind of memory in gravitationally-driven evolution is widely acknowledged and described, at cosmological scales \citep{crocce2006} and intergalactic scales \citep{zaroubi96,aubert2004,knebe2008,libeskind2012}. At galactic scales this memory is at the basis of galactic archeology \citep{helmi1999}.
Therefore, is it possible that some correlation between reionization history and position within the MW survives until today?
In this paper we set to answer this question by analysing simulations of the reionization of the local group in terms of the relation between a satellite's reionization history and its position within the MW halo at z=0. Moreover, the present study brings an important improvement in our radiative transfer scheme, which now includes a simple recipe for radiative feedback, in the spirit of \cite{iliev2007}. 
The paper is laid out as follows: first we describe the simulation used and radiative transfer postprocessing technique (Sec. \ref{s:methodology}). We then proceed to our results (Sec. \ref{s:results}), and discuss them (Sec. \ref{s:discussion}), before presenting our conclusions.

\section{Methodology}
\label{s:methodology}

The methodology of this study is similar to paper I but features an improved radiative transfer scheme, accounting for radiative feedback suppressing star formation in low mass haloes.

\begin{table*}
  \include{tab4paper2}
  \label{t:models}
\caption{Properties of the models used. Column (2) gives the criterions of the source models, used to mimic Lyman-Werner and supernova feedback. Column (3) takes the value ``yes'' for models with radiatively regulated star formation, and ``no'' otherwise. Column (4) gives the halo emissivity per solar mass of dark matter halo, except for the SPH model where the emissivity is given per mass of young stars ($<30$ Myr, i.e. the duration between 2 snapshots of the SPH simulation), hence the $^{\star}$ superscript. In all cases, the emissivity is given {\em after} accounting for an escape fraction $\fesc=0.2$. Column (5) gives the reionization redshift of the MW and M31 progenitors for each model, i.e. the time when the mass-weighted ionized fraction of the progenitor reaches 0.5. Column (6) gives the duration of the progenitors' reionization as the time spent to increase the mass-weighted ionized fraction $\xm$ from 0.1 to 0.9. Column (7) gives this duration in Myr. The 4 first models are our baseline models,a and the 5th is taken from paper I.}
\end{table*}

\subsection{The CLUES simulation}
\label{s:simclues}
The simulation used in this work was performed in the framework of the CLUES project \citep{clues2010}\footnote{\url{http://www.clues-project.org}, seed number 186592}. 
It was run using standard Lambda cold dark matter ($\Lambda$CDM) initial conditions assuming a WMAP3 cosmology \citep{spergel2007}, i.e. $\Omega_{\rm{m}}=0.24$, $\Omega_{\rm{b}}=0.042$, $\Omega_{\Lambda}=0.76$. A power spectrum with a normalization of $\sigma_8=0.73$ and $n=0.95$ slope was used. The PMTree-SPH MPI code GADGET2 \citep{springel2005} was used to simulate the evolution of a cosmological box with a side length of $64$ \hmpc. Within this box a model Local Group that closely resembles the real Local Group was identified using a 1024$^3$ particles run (cf. \cite{libeskind2010}). This Local Group was then re-sampled with 64 times higher mass resolution in a region of 2 \hmpc about its center giving an equivalent resolution of 4096$^3$ particles, i.e. a mass resolution of $\mdm=2.1 \times 10^5 h^{-1} \Msun$ for the dark matter and $m_{gas}=4.42 \times 10^4 h^{-1} \Msun$ for the gas particles. For more detail we refer the reader to \cite{clues2010}. The feedback and star formation prescriptions of \cite{springel2003} were used. Outputs are written on average every 30 Myr.
The simulation starts at z=50. As it runs, dark matter and gas collapse into sheets and filaments, extending between halos, as comprehensively described in \cite{ocvirk2008,codis2012,hoffman2012,libeskind2012}. These feed proto-galaxies which then start forming stars.
It includes a uniform rising UV cosmic background generated from quasi-stellar objects and active galactic nuclei \cite{haardtmadau96}, switched on at z=6. Therefore the radiative transfer computations that we perform will be valid only at earlier times. We will see that this is not a problem, since our models always achieve complete reionization before z=6.
This simulation has been used to investigate a number of properties of galaxy formation at high resolution \citep{jaime2011,knebe2011b,knebe2011a,libeskind2011a,libeskind2011b}. Besides being a well-studied simulation, the advantage of this dataset for the present study is twofold. First of all, it produces a fairly realistic local group at z=0: the MW and M31 are in the correct range of masses and separation. Secondly, its mass resolution in the zoomed region allows us to resolve the $10^7 h^{-1}\Msun$ haloes. This is of crucial importance in reionization studies since they are the most numerous sources of UV photons.

\subsection{Radiative post-processing}
\subsubsection{ATON}
ATON is a post-processing code that relies on a moment-based description of
the radiative transfer equations and tracks the out-of-equilibrium ionisations
and cooling processes involving atomic hydrogen \citep{aubert2008}. Radiative quantities (energy
density, flux and pressure) are described on a fixed grid and evolved
according to an explicit scheme under the constraint of a Courant-Friedrich-Lewy condition (hereafter CFL). The simulations presented in this work used a mono-frequency treatment of the
radiation with a typical frequency of 20.27 eV for 50000 K black body
spectrum. Because of the high resolution of the CLUES simulation, we do not make any correction for the clumping, as was done for the largest boxes of \cite{aubert2010}. ATON has been ported on multi-GPU architecture, where each GPU handles a cartesian sub-domain and communications are dealt using the MPI
protocol \citep{aubert2010}. By achieving an x80 acceleration factor compared to CPUs, the CFL
condition is satisfied at high resolution within short wallclock computing
times. As a consequence, no reduced speed of light approximation is necessary
and it may be of great importance for the timing arguments of the local
reionization discussed hereafter.
Along the course of this work, simulations were run on segments of 8 to
64 GPUs on the Titane and Curie machines of the CCRT/CEA supercomputing facility, with typically 160000 radiative timesteps performed in 37 hours.

The postprocessing approximation has potentially important consequences on our results, as discussed for instance in \cite{baek2009,frank2012}. While the temperature of the gas is consistently followed by ATON, the gas density is ``frozen'' to that given by the SPH simulation snapshots. 
This means that our scheme does not allow for photo-evaporation, however we do include a simple recipe for the effect of photo-heating of the baryons in haloes, resulting in suppression of star formation in low-mass haloes, in the spirit of \cite{iliev2007}. More details are given in Sec. \ref{s:FB}.
By design, self-shielding is also accounted for, and results in a later reionization of sourceless high density regions, such as mini-haloes or the cold gas filaments. 

\subsubsection{Fields setup}
The gas density field is projected onto a $512^3$ grid of 11 comoving \hmpc side.
The center of the grid is the barycenter of all the particles which end up within 300 \hmkpc of the MW at z=0. 
This setup gives us a spatial resolution of ${\Delta x} = 21 \, {\rm{ h^{-1} kpc}}$. The sources are projected on the same grid. 
As explained in Sec. \ref{s:simclues}, the CLUES simulation uses a zoom technique, with a high and low resolution domains. The high resolution (hereafter HR) domain contains the objects of interest (MW and M31), and is described with dark matter, gas and star particles. At $512^3$ resolution all grid cells contain at least one gas particle in the HR region in the highest redshift snapshot (z=19.5). On the other hand, the low resolution (hereafter LR) domain does not have any SPH particle. Therefore we set the gas density in the low resolution domain to $\rho_{\rm LR}= 10^{-2} \rho_{\rm C}$, where $\rho_{\rm C}$ is the critical density of the Universe. The LR region does not contain any stars either. Photons reaching the HR/LR boundary region just  leave the local group and quickly reach the edges of the computational box. There, we use transmissive boundary conditions, i.e. light just exits the box.

\subsubsection{Ionizing sources}
\label{s:sources}
Our model is based on dark matter haloes catalogues produced using the Amiga halo finder\footnote{\url{http://popia.ft.uam.es/AHF}} \citep{AHF04,AHF}. 
We keep only the haloes which have 100\% of their mass in high resolution  dark matter particles. Dwarf galaxies of the early Universe  are subject to a wide range of feedback processes beyond photo-evaporation by a UV background. Although our code does not allow for live self-regulation of the sources, we tried to account for the influence of at least some of the relevant feedback processes.
We use a constant $\fesc=0.2$, which is among values allowed by recent studies on the UV continuum escape fraction of high-z galaxies \citep{wise2009,razoumov2010,yajima2011,wise2014}. We neglect any possible AGN-phase of our emitters. Such sources could already be in place in rare massive proto-clusters during reionization \citep{dubois2011,dubois2012}, and contribute to the cosmic budget of ionizing photons \citep{haardt2011}, but they are beyond the scope of the present study.
The properties of our source models are summarized in Tab. \ref{t:models}.


As in paper I, we consider that all haloes with a virial temperature ${\rm T_{vir}<10^4 K}$ are unable to form stars due to the Lyman-Werner background dissociating H$_2$, the only coolant of pristine hydrogen gas at these masses \citep{barkana2001,ahn2009}. More details are given in paper I. In essence, we consider only haloes with ${\rm T_{vir}>10^4 K}$ as UV sources. 

We assign an instantaneous star formation rate to each halo, assuming SFR $\propto  M$. However since the total emissivity of a given halo depends on its mass times the star formation efficiency times the emissivity times the escape fraction and is therefore degenerate with respect to the last 3 parameters, it suffices to set the global emissivity of our models.

We also re-use two models of paper I: SPH and H44 (SPH and H44 SNfb in Tab. \ref{t:models}) in order to investigate the impact of an alternative source modelling (see Sec. \ref{s:rrmaps}) and a case of external reionization of the MW by M31 (see Sec. \ref{s:external}).

A summary of the properties of our models is given in Tab. \ref{t:models}.
The emissivities are given in ${\rm photons/s}$ per $\hmo$ $\Msun$ of dark matter halo and per $\hmo$ $\Msun$ of young stars ($<30$ Myr). While this duration is larger than the typical 10 Myr used for the lifetime of massive, UV-bright stars, it is imposed by the temporal spacing between 2 snapshots of the SPH simulation which produced the star particles. This is equivalent to smoothing the star formation history of the simulation over 30 Myr, and this is not likely to affect our results.

The adopted emissivites give reionization redshifts for the local group galaxies progenitors between 8 and 13, i.e. well within the range allowed by observations \cite{fan2006} and large-scale simulations such as \cite{alvarez2009}.

\subsubsection{Simple radiative feedback}
\label{s:FB}
\begin{figure*}
  {\includegraphics[width=0.33\linewidth,clip]{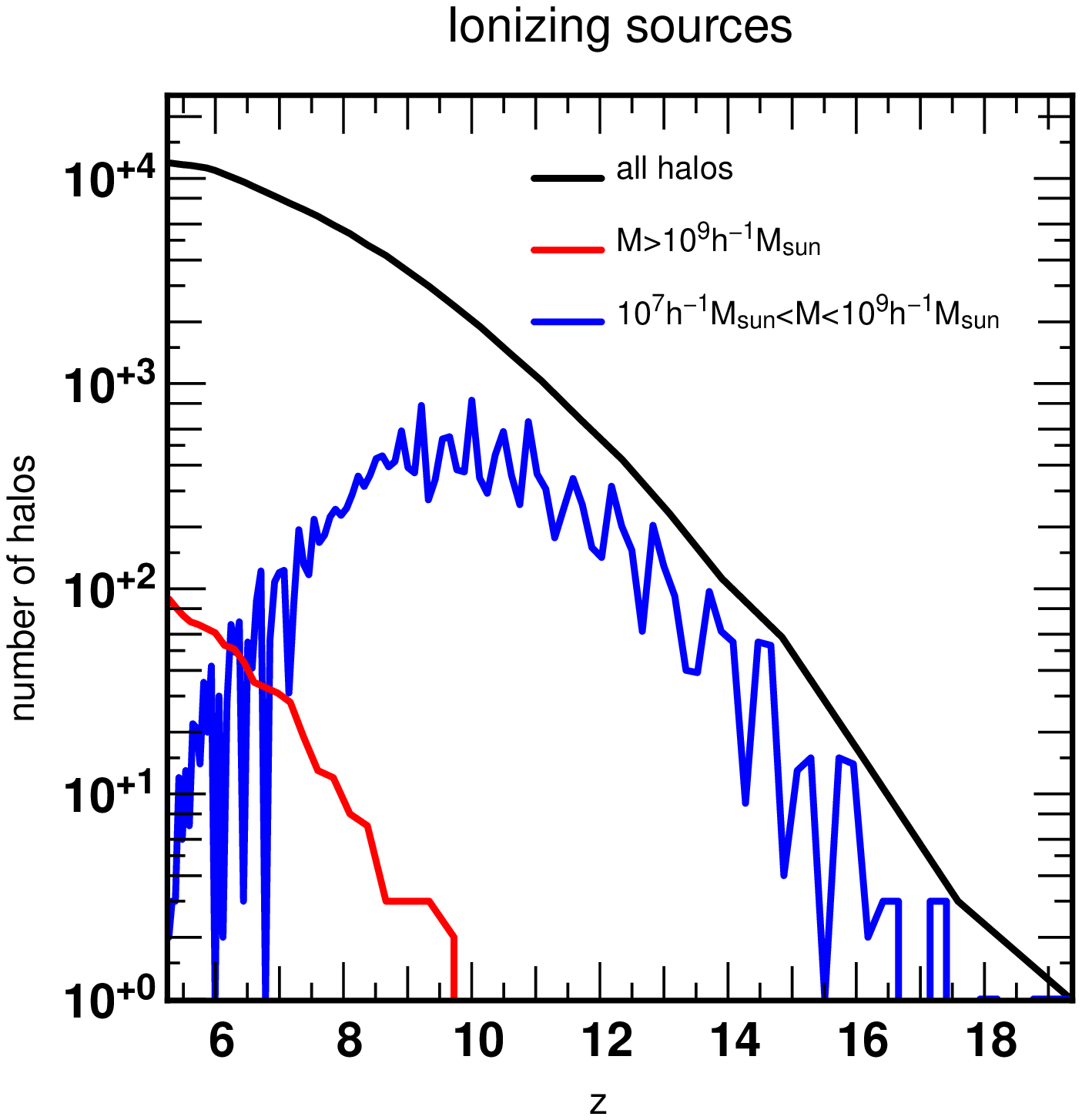}}
  {\includegraphics[width=0.33\linewidth,clip]{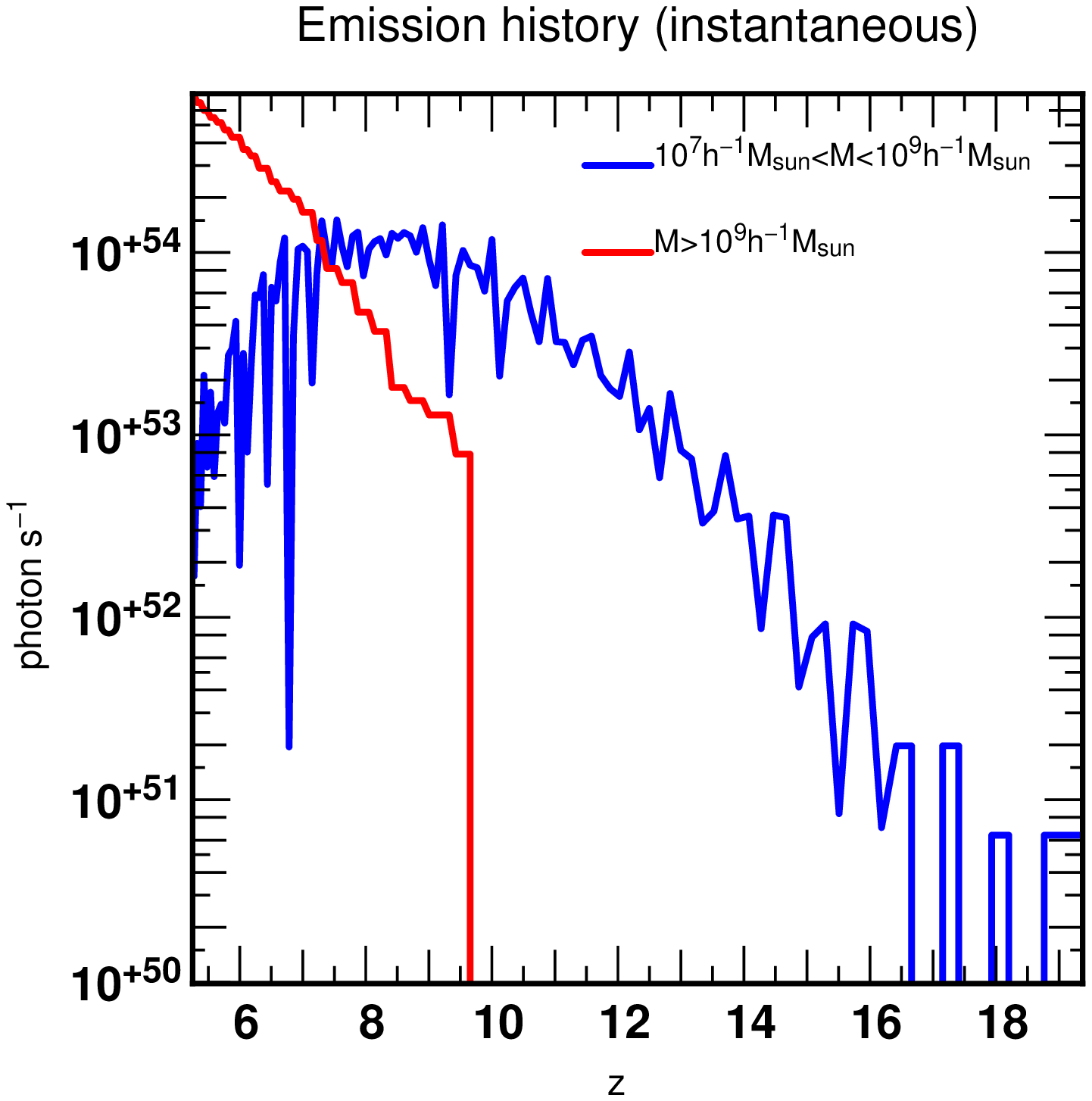}}
  {\includegraphics[width=0.33\linewidth,clip]{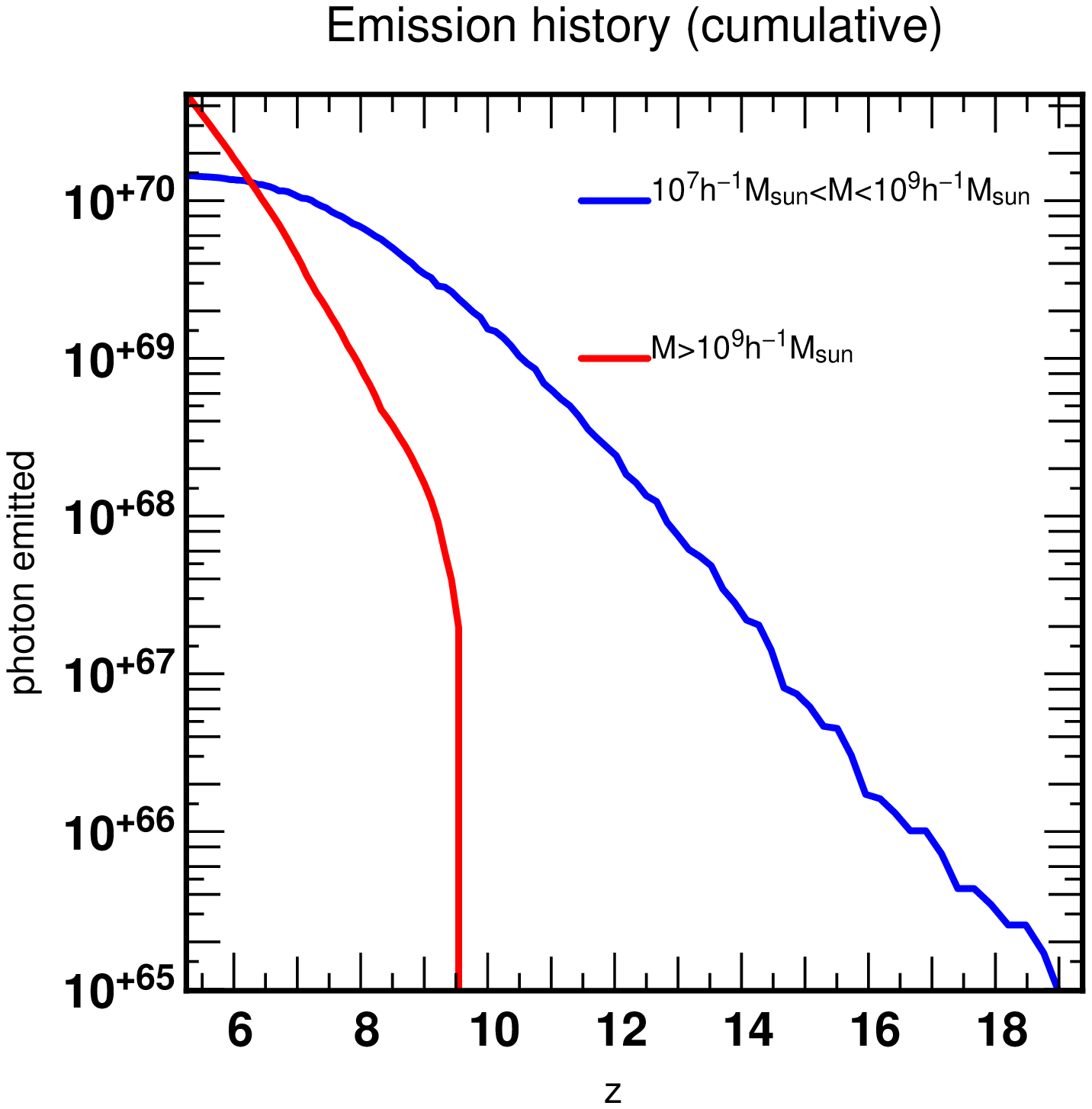}}
  \caption{Illustration of the radiative feedback model for the sources. {\em Left:} number of ionizing sources in the box. {\em Middle:} Total emissivity of the low-mass and high-mass haloes with time. {\em Right:} Time-integrated emission history of the low mass and high mass haloes.}
\label{f:nicfb}
\end{figure*}

In a fully coupled radiative-hydrodynamics simulation, the gas field reacts to the photo-heating, and can result in the dispersion of low-mass gas structures \citep{shapiro2004,iliev2005,iliev2009}. This should induce a form of self-regulation of star formation and therefore emissivity by shutting off sources in the ionized low-mass haloes, as shown in \cite{iliev2007}. Even though a small number of coupled galaxy formation codes have recently been built \citep{petkova2011,rosdahl2011,finlator2011,enzomoray}, at the moment no application to the formation of the local group in a zoom simulation such as the CLUES dataset we use here has been performed, mainly because of the huge computational cost involved. 

In order to account for this effect in our semi-analytical model, we divide our sources into 2 classes:
\begin{itemize}
\item{low-mass, radiatively regulated haloes: these are massive enough not to be suppressed by LW background, i.e. they are atomically cooling haloes (${\rm T_{vir}>10^4 K}$), but are still sensitive to photo-heating. Using coupled radiative-hydrodynamical numerical simulations, \cite{pawlik2013} showed that above $10^9 \Msun$, a proto-galaxy is able to self-shield and is dense enough not to be affected by radiative feedback any more. We use this value as the upper mass of the low-mass, radiatively regulated haloes class. These haloes are UV sources only when the ionized fraction of their cell is smaller than $\xsupp=0.5$. In this ``neutral'' state, they will form stars at some prescribed rate, which will live for 10 Myr. As a result, they undergo radiatively-driven cycles of star formation - photo-heating - suppressed star formation - cooling $+$ recombination - star formation and therefore can turn on and off several times through the course of the simulation. This results in the oscillations seen in Fig. \ref{f:nicfb}.}
\item{high-mass, self-shielding star forming haloes: these objects are massive enough ($>10^9 \Msun$) to keep forming stars even when their cell is completely ionized. We know from high-resolution simulations such as \cite{pawlik2013} that such haloes are able to self-shield, but we do not have the spatial resolution in our RT simulation to resolve the shielded region. Therefore we consider that such massive haloes will continue forming stars even if the RT cell in which they reside is completely ionized. Therefore these haloes are UV-bright throughout the simulation.}
\end{itemize}

The basic behaviour of this model is illustrated in Fig. \ref{f:nicfb} for a H7e43 UVfb model. For this example only, a $256^3$ grid was used instead of $512^3$, but the basic behaviour is identical to the higher resolution $512^3$ runs of the rest of the paper. In this example the box reionizes at $z \sim 6$. The left panel shows the number of ionizing sources. The blue line shows the number of active low-mass haloes. At high redshift, it is close to the total number of haloes (low and high masses), but radiative self-regulation suppresses a fraction of the low mass haloes, giving rise to the oscillations. The volume-filling fraction of these haloes and their Stromgren spheres increases rapidly with decreasing redshift, and at $z \sim 12$, some of them start to overlap, leading to an increasing offset between the maximum of emitting low mass haloes and the total number of haloes (i.e. a smaller and smaller fraction of the low mass haloes is a source). This becomes even more marked by $z \sim 10$, where the first high mass haloes, immune to radiative feedback, start to appear. Their Stromgren spheres rapidly expand, suppressing star formation in nearby low mass haloes until the latter are almost totally shut down. The high-mass sources finally outnumber the low-mass sources at $z \sim 6$. However, due to their high mass and therefore larger photon output (and considering a constant $\fesc$ for all halo masses), they outshine the low-mass haloes as early as z$\sim 7$, as shown by the middle panel of Fig. \ref{f:nicfb}.
This evolution is similar to that reported in \cite{iliev2007,wise2014}, where low-mass galaxies are responsible for initiating  reionization at high redshift, but are gradually radiatively suppressed by the more massive haloes insensitive to the UV background which finally outshine them. However since they appear earlier, the cumulated photon output of the low mass haloes through cosmic history is actually comparable to that of the massive haloes at the time of overlap, although the Universe is kept ionized by the high-mass haloes afterwards. The time at which the cumulated photon output of the 2 classes of haloes are equal actually depends on source emissivity. 
The suppression of low-mass sources reduces the overall emission of UV photons through cosmic time. As a result, models with UV feedback (``yes'' in column (3) of Tab. \ref{t:models}) should reionize slower than models without UV feedback. This is confirmed for instance by the reionization timings (columns 5,6,7 of Tab. \ref{t:models}) of model H7e43 UVfb versus model H7e43 NOfb (same emissivity but without feedback, therefore more sources active at any given time). In the former the MW reionizes at z=10.5, while in the latter, it reionizes at z=11. The same holds for M31.

\section{Results}
\label{s:results}
\begin{figure*}
\begin{tabular}{cc}
  {\includegraphics[width=0.49\linewidth,clip]{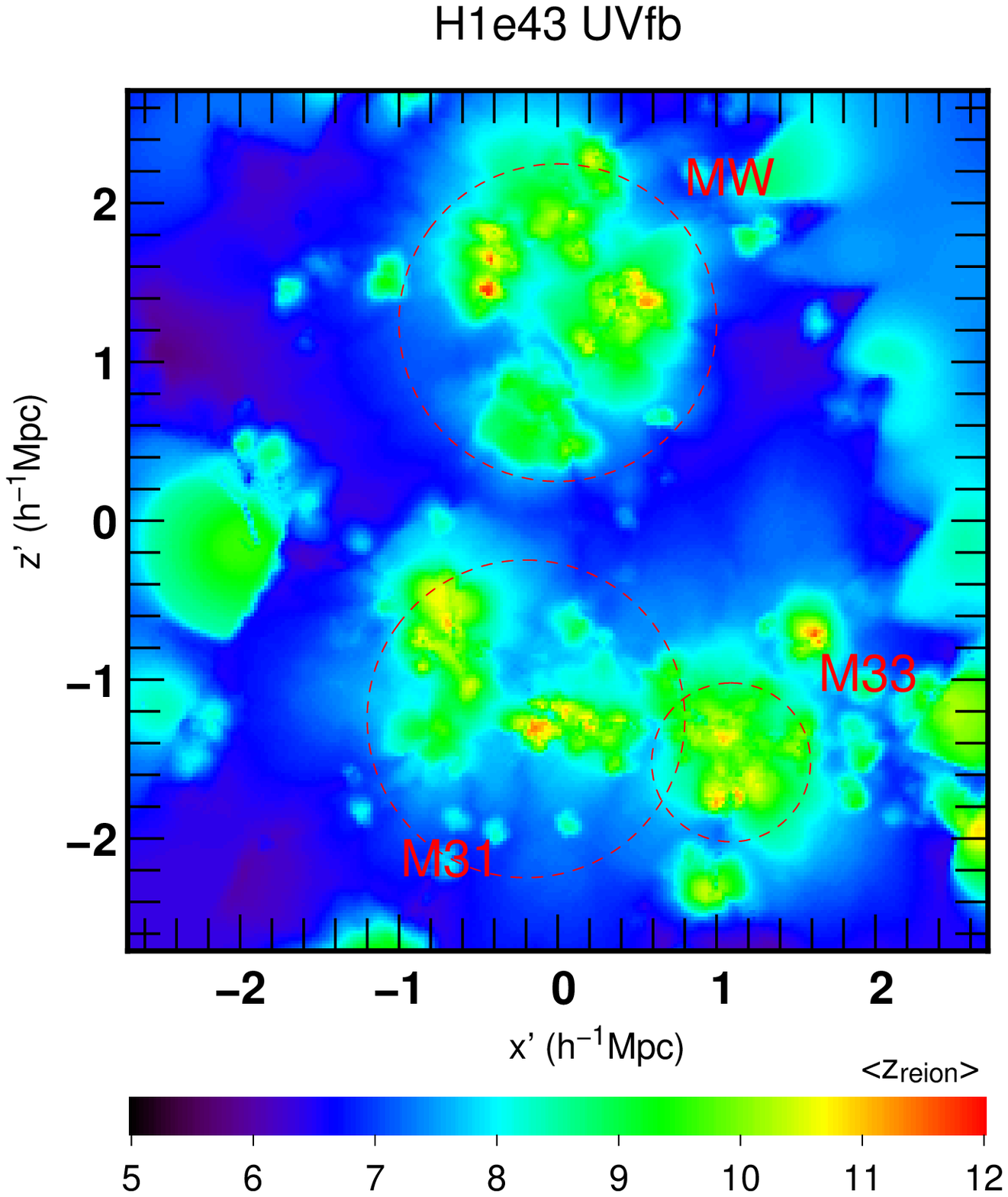}}&
  {\includegraphics[width=0.49\linewidth,clip]{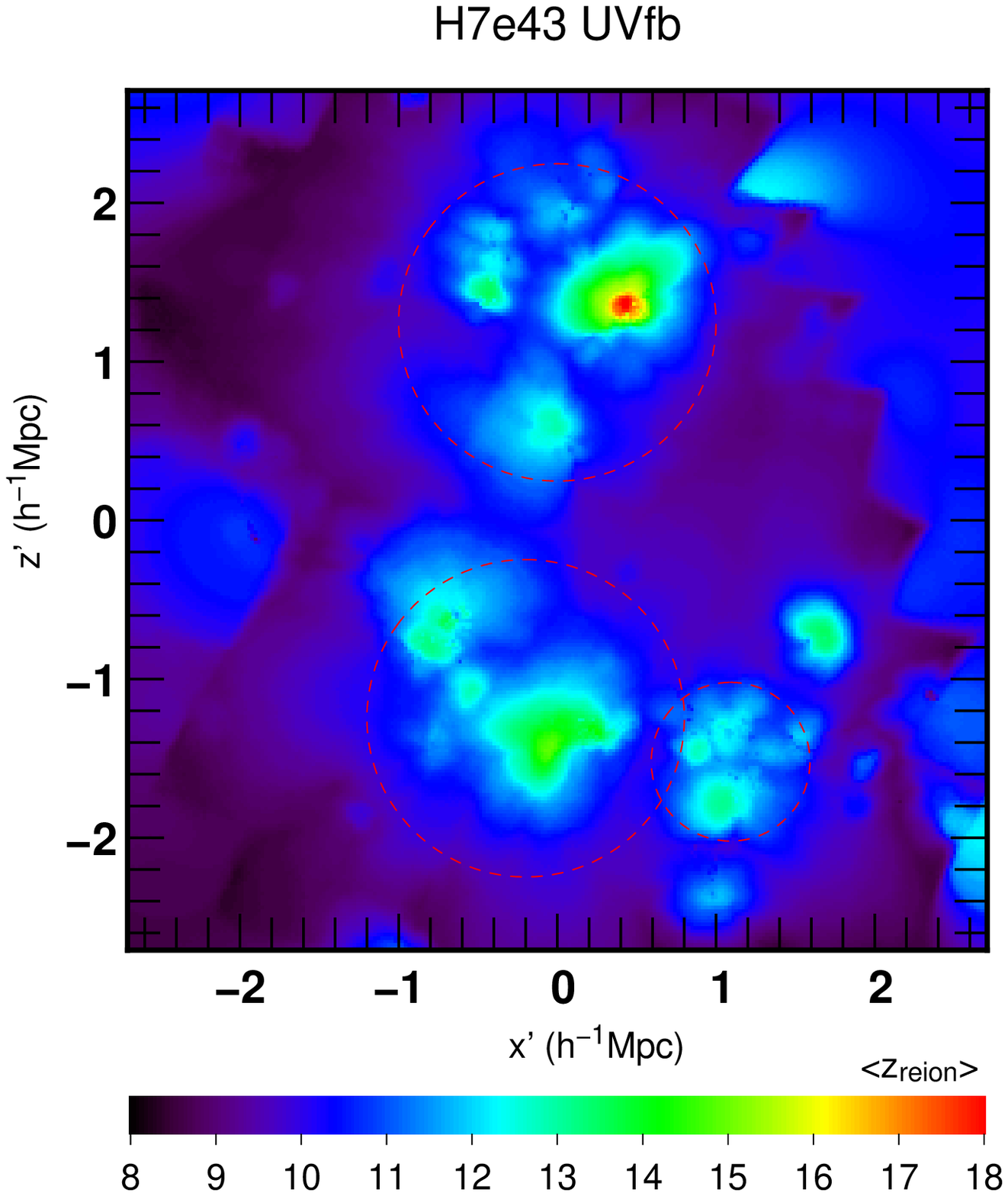}}\\
  {\includegraphics[width=0.49\linewidth,clip]{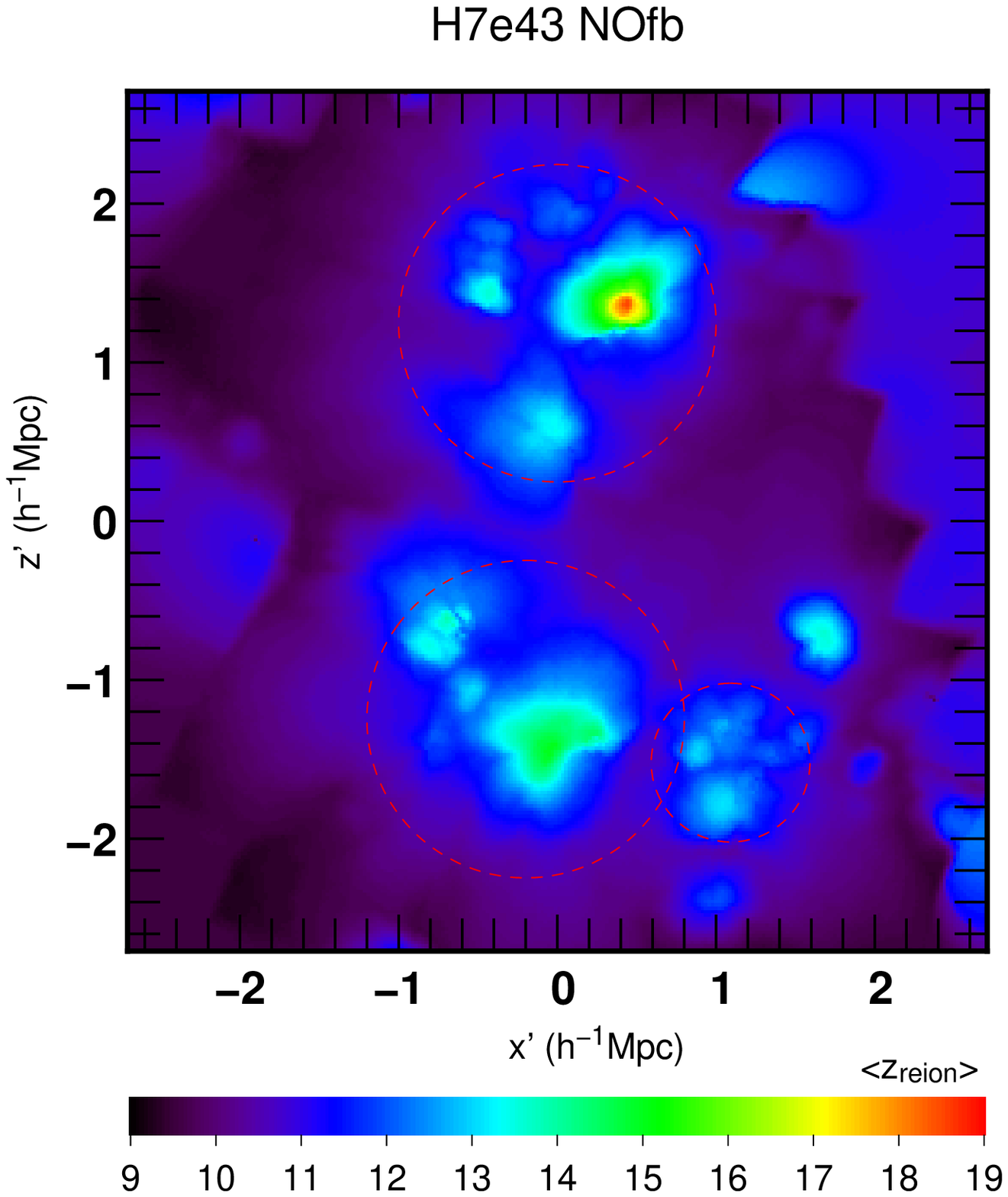}}&
  {\includegraphics[width=0.49\linewidth,clip]{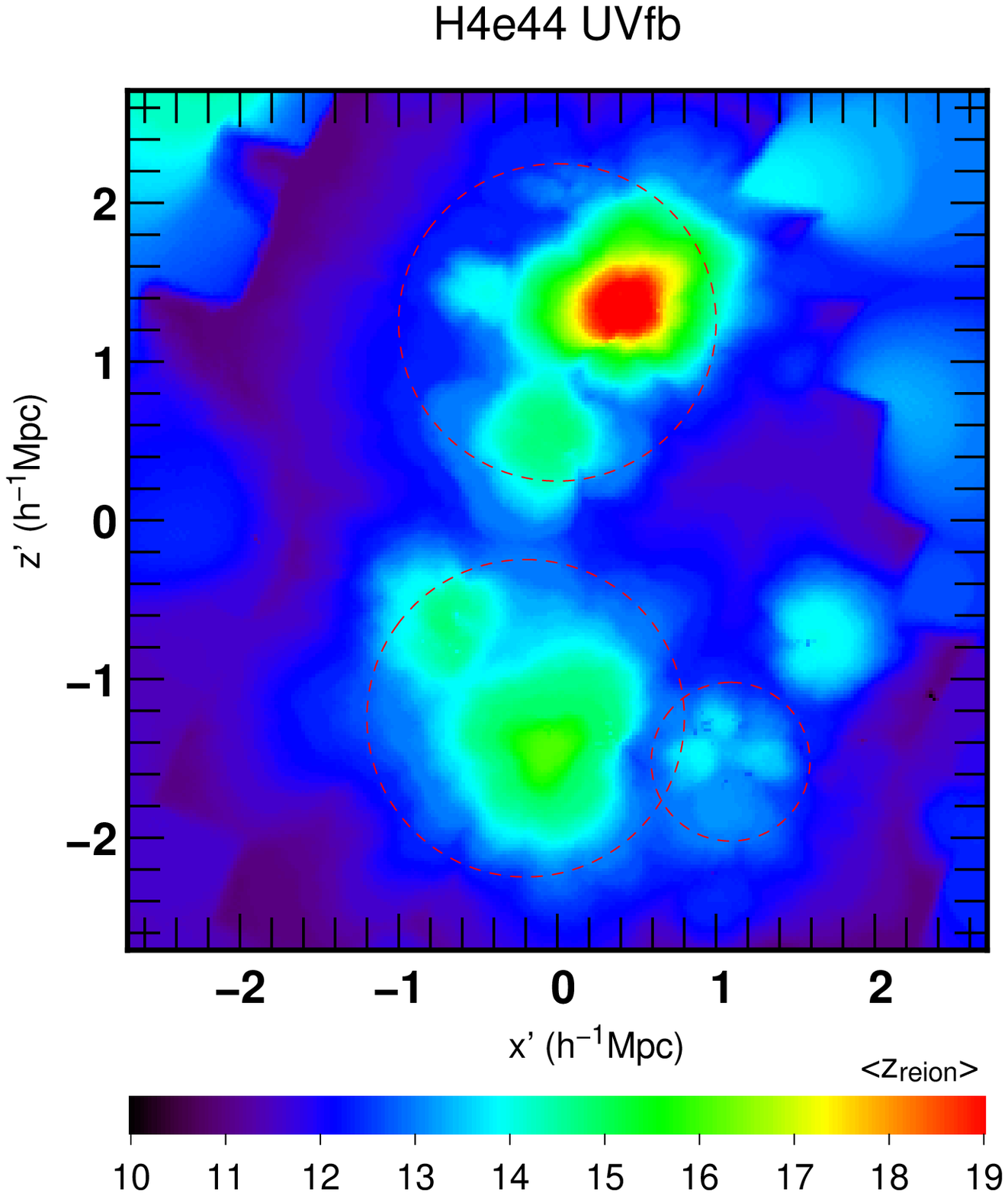}}
\end{tabular}
  \caption{Reionization maps of the local group of galaxies for our 4 baseline models. The simulation domain has been cut in the plane containing the centers of M31, M33 and the MW. The maps are obtained as the average $\zreion$ of a slab of 0.2 \hmpc thickness centered on this plane. The color codes the reionization redshift of each cell. The red dashed circles show the approximate location of the progenitors of the 3 major local group galaxies (Top middle: MW, bottom middle: M31, bottom right: M33). The square artifacts in the corners are due to the transition from the high to low resolution domains of the SPH simulation.}
\label{f:rmaps}
\end{figure*}

In this section we check the good behaviour of our RT scheme with radiative feedback using classical eulerian reionization maps. Then we introduce the lagrangian reionization map technique, using dark matter haloes as tracers. We then turn to computing the reionization histories of the satellite populations of MW and M31 and show that they also depend on satellite positions.

In general the eulerian reionization maps we obtain are rather similar to those of paper I, however with slightly more small-scale structure at a given emissivity. 
For instance, the low emissivity scenario H1e43 UVfb has much more small-scale structure than the low emissivity scenario of paper I. In particular there are a number of small early-reionized bubbles ($100-200$ \hmkpc across at some distance from the main patches of each progenitor, as well as outside of the progenitors. These are produced by small self-regulated haloes in low-density IGM: they turn on, ionized their neighbourhood, self-suppress, therefore turning off. However the density of the IGM in these regions is too low for the gas to recombine and a unique episode of star formation in these low mass haloes is enough to produce these small patches in the map.
The H7e43 UVfb and H7e43 NOfb allow us illustrate the impact of the feedback recipe on the reionization map: again, the maps are very similar, except that the model with UV feedback reionizes slightly later, and has more small scale structure. This is not surprising: feedback reduces overall emissivity, and paper I showed that reducing emissivity leads to later reionization and more small scale structure. 
Moreover, the H4e44 UVfb model, with the highest emissivity, also shows the smallest degree of small scale structure, in agreement with paper I. 
This validates our understanding of the behaviour of our model with feedback.
Finally, we see that the addition of UV feedback does not change the basic conclusion of paper I, which is that the MW and M31 appear to reionize in isolation within the emissivities considered here.

\subsection{Lagrangian reionization maps}

\label{s:parts}
\begin{figure*}
\begin{tabular}{cc}
{\includegraphics[width=0.47\linewidth,clip]{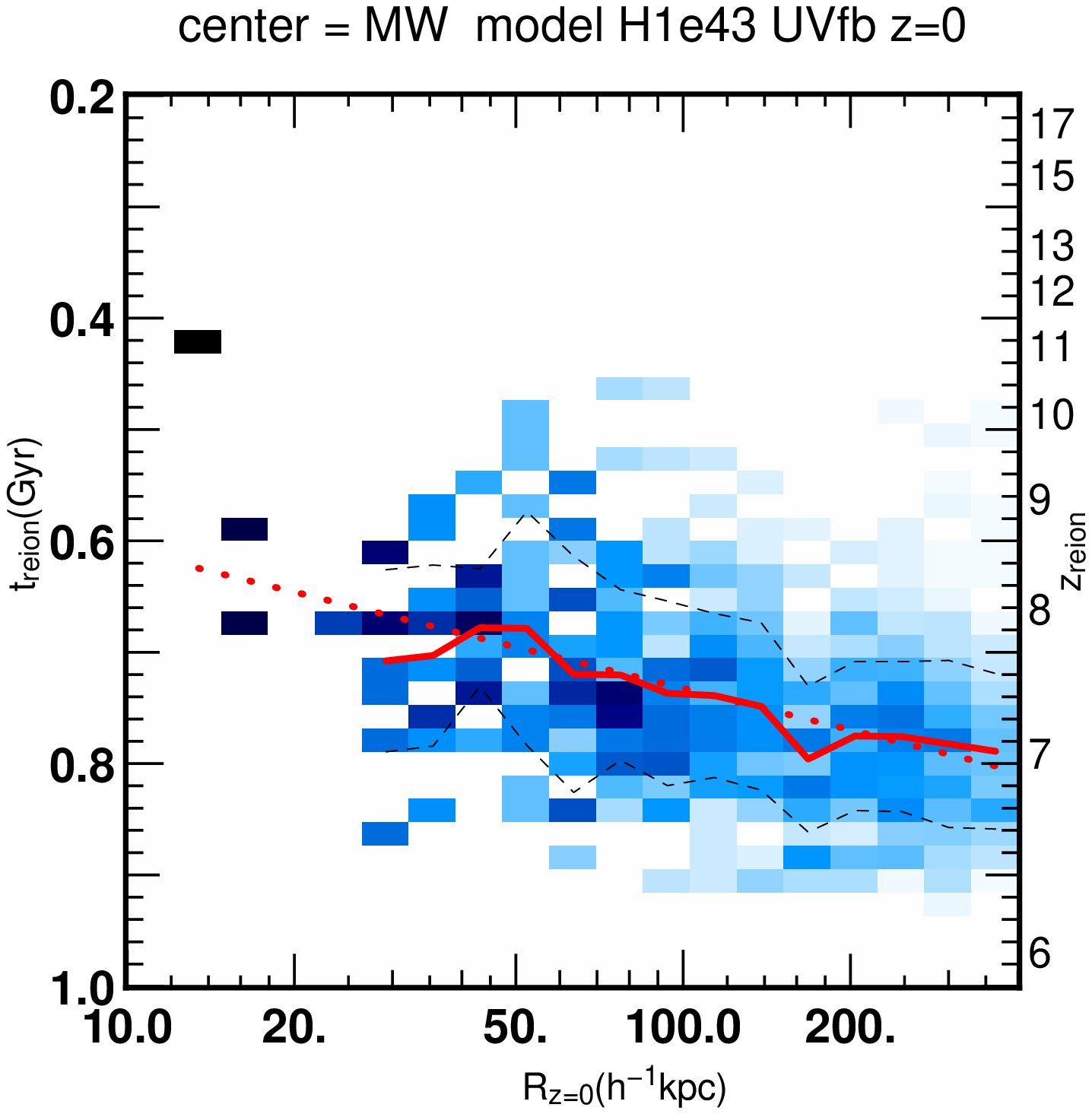}} &
  {\includegraphics[width=0.47\linewidth,clip]{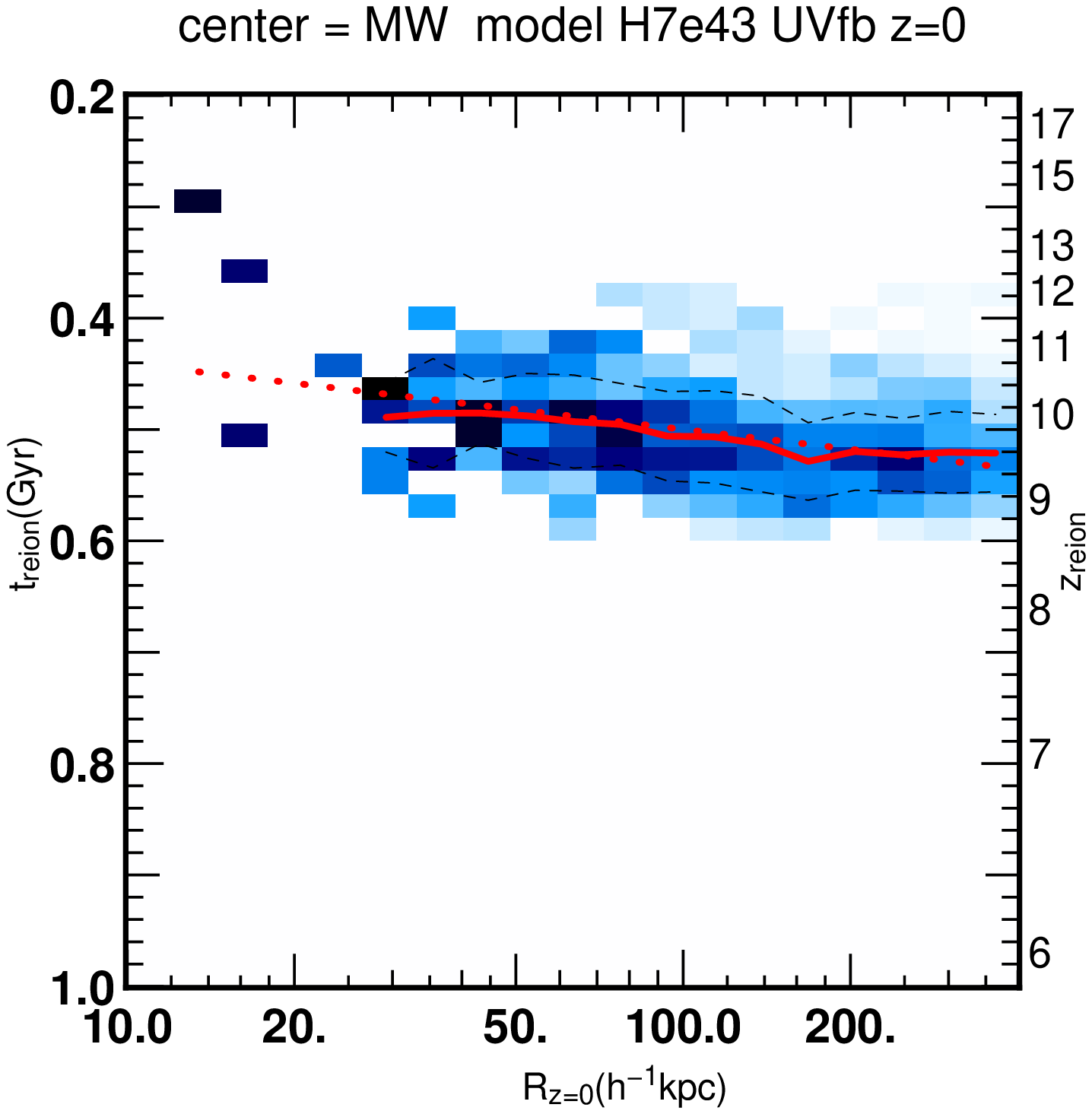}} \\
{\includegraphics[width=0.47\linewidth,clip]{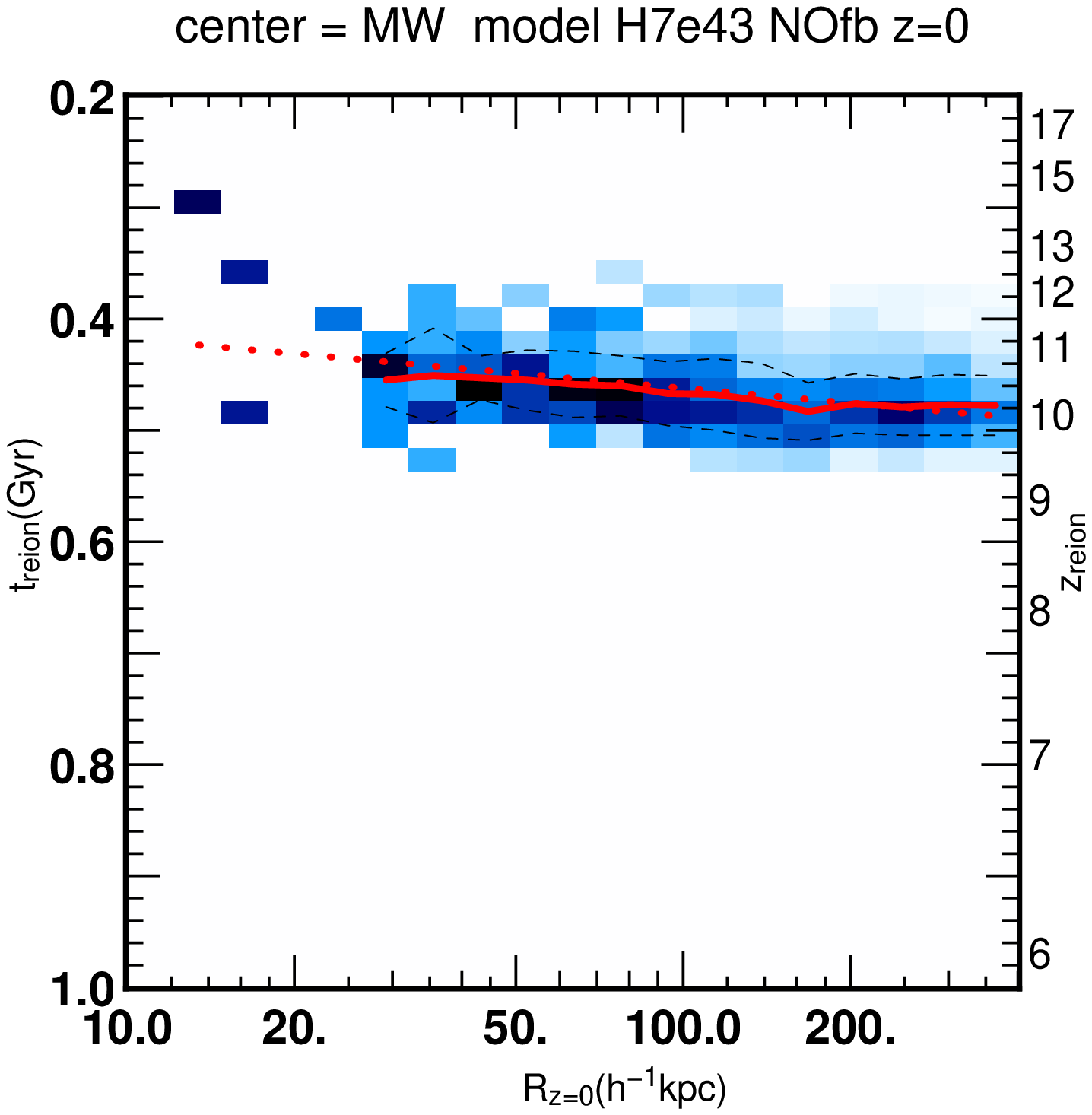}} &
  {\includegraphics[width=0.47\linewidth,clip]{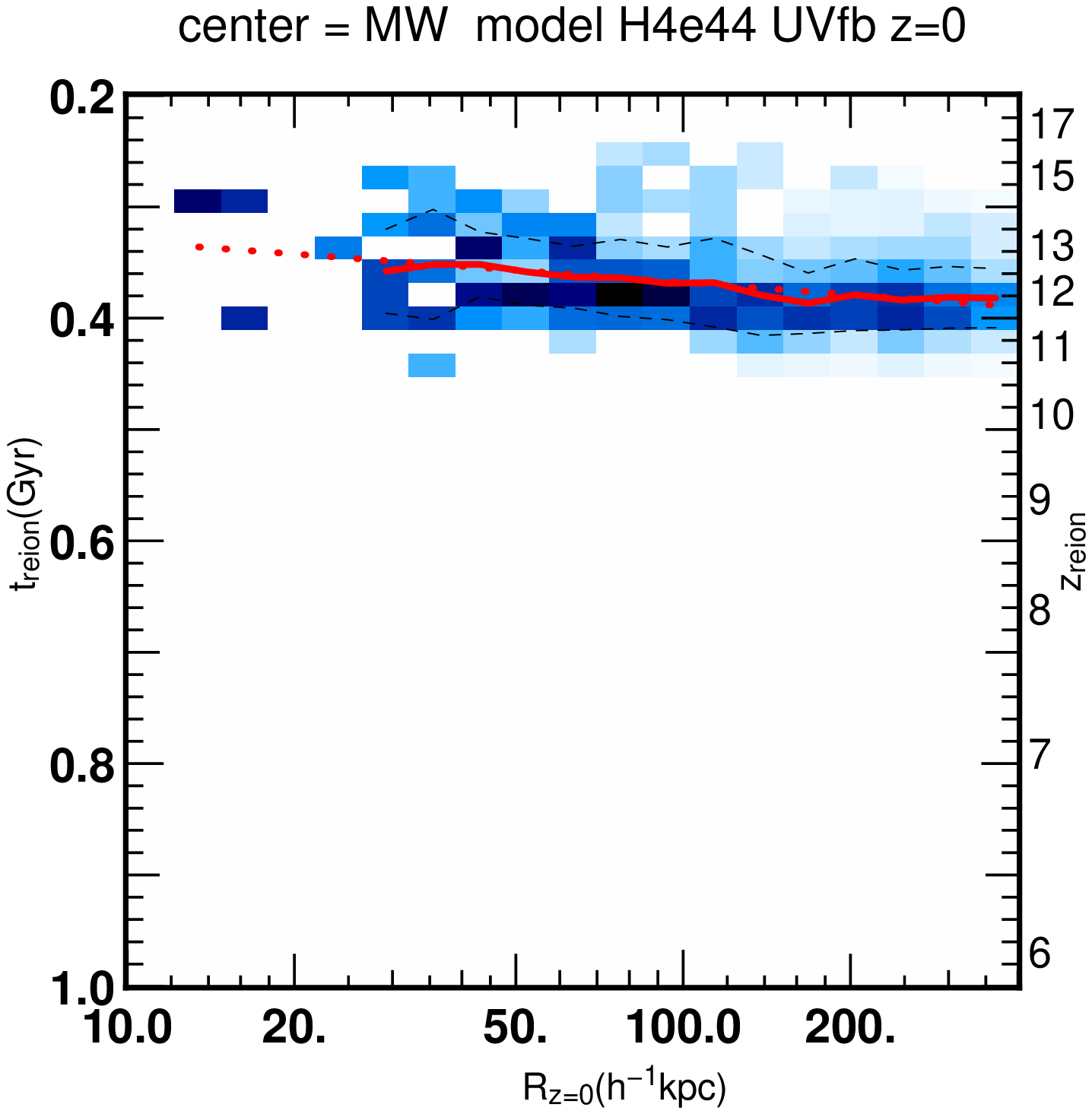}} \\
\end{tabular}
\caption{Distribution of the reionization redshifts of the MW satellite halos as a function of position at z=0. The color codes the relative number of of satellites in each cell, with each radial bin normalized by a fit to the number of satellites per bin. For instance, the darkest inner cells contain only 1 satellite each. The red solid line shows the average satellite $\zreion$ for each radial bin, and the black thin dashed lines the dispersion of $\zreion$. The red dotted line shows the linear fit to the average, with parameters given in Tab. \ref{t:res}.}
\label{f:rmaplog}
\end{figure*}

\subsubsection{Dark matter haloes and particles as tracers}
In order to investigate the evolution and possible survival of the structures seen in the reionization maps during the progenitor's collapse and galaxy assembly, we build lagrangian reionization maps.
Contrary to the eulerian reionization maps, which are grid-based, lagrangian reionization maps use the dark matter haloes or particles of the simulation as tracers. 
In each snapshot, we assign to the particles the $\xion$ of the cell that particle resides in. This yields for each particle a reionization history, from which we define a reionization redshift for each particle as the last redshift where $\xion < 0.5$. 
Since we are interested in understanding the impact of patchy reionization on the satellite population, we select haloes with a mass between $10^7-10^9 \, \hmo \Msun$ at z=0, i.e. the mass range expected for ultra-faint dwarfs halos \citep{ocvirk2011}. This low-mass satellite population is supposed to be the most sensitive to photo-heating. More massive satellites, such as LMC or SMC, are not affected by reionisation since they are still forming stars today. This is why we restrict ourselves to the low-mass satellite population. Note that we only consider the surviving satellites at z=0 and not the disrupted ones. We compute reionization redshift of a z=0 dark matter halo as the average reionization redshift of its core dark matter particles ($<0.1 \Rvir$), so as to minimize contamination by background particles.
We can then produce lagrangian reionization maps at any redshift. The main novelty of such maps is that they account by construction for the dynamical evolution of the system down to the epoch of interest. In particular, unlike eulerian reionization maps, they allow us to investigate the structure of the reionization history of the MW at z=0. 

\subsubsection{Radial reionization maps}
\label{s:rrmaps}

We computed the reionization redshifts of the $z=0$ surviving dark matter haloes of the MW and M31 for our 4 baseline models. We then investigate the relation between the haloes positions at $z=0$ and their reionization redshifts ($\zreion$) and reionization times ($\treion$, with the Big Bang as origin $\treion=0$), by means of the $\zreion$ - distance to the MW center distribution shown in Fig. \ref{f:rmaplog}.

\paragraph{Inside-out reionization of the satellite population}
Let us for this paragraph consider the H1e43 UVfb model (top left panel of Fig. \ref{f:rmaplog}). Most of the MW sub-haloes reionize between $z=6$ and $z=10$. Although there is a large dispersion within these values (the black dashed lines show the dispersion of the distribution), there is a clear inside-out gradient in $\zreion$, shown by the thick red line (the average of the distribution in each radius bin). 
 The average reionization redshift of the satellites is clearly correlated with their position at z=0: even today, the inner satellites show a tendency to reionize earlier than their outer halo counterparts. The dynamical evolution of the MW system since reionization more than 12 billion years ago, has partially conserved the relative radial positions of the satellite progenitors, in such a way that the reionization of the satellite system still appears inside-out at z=0. We show the $\zreion$ - galactocentric distance distribution out to 400 $\hmo$ kpc only (i.e. 570 kpc) so as to avoid contamination by M31 sub-haloes (the 2 galaxies are separated by 888 $\hmo$ kpc, i.e. 1268 kpc in this simulation). 
The gradient in average $\zreion$ or $\treion$ is surprisingly steady. We find that the mean reionization time can be well approximated by the following law:
\begin{equation}
\treion = \treion^{1 {\rm kpc}} - \frac{{\Delta} \treion}{{\Delta} \log R} \log R \, ,
\label{eq:linfit}
\end{equation}
where $\treion^{1 {\rm kpc}}$ is the average reionization time at 1 kpc from the Galaxy center, and  ${\Delta}\langle \treion \rangle /\Delta\log R$ is the slope of the gradient between 1 and 400 \hmkpc. 
The center of the MW reionized on average about 200 Myr earlier than the outskirts, and the $\treion$ gradient has a slope ${\Delta}\langle \treion \rangle /\Delta\log(R) \sim 123$ Myr/dex.
The same formalism holds for the reionization redshifts. The values of the fit parameters are given in Tab. \ref{t:res} for the 5 scenarios. The linear fit is shown by the red dotted line. It gives an accurate approximation of the gradient in all cases, and thanks to the weighting applied (in square root of the number of objects in the radius bin), the fit is effectively rooted in the bulk of the satellite system, and not biased by the few rare innermost haloes. 

\begin{table*}
\label{t:res}
\include{tab4papergrads}

\caption{Properties of the $\zr$ gradient of surviving haloes at z=0, measured in MW and M31 for our 5 models. For the SPH model the emissivity is given per mass of young stars hence the $^{\star}$ superscript. Columns (4-5) are related to the linear fit to the average reionization redshift or time shown in the halo lagrangian reionization maps. Column (5) gives the slope of the reionization redshift gradient, while column (4) gives the central value. These  are explicited in Eq. \ref{eq:linfit}. Column (6) gives the dispersion of reionization redshifts around the fit.}
\end{table*}

\paragraph{Impact of emissivity}
Increasing emissivity reduces the dispersion of $\zreion$ and $\treion$ and more importantly flattens the radial $\treion$ gradient, as shows a quick comparison between the H1e43 UVfb, H7e43UVfb and H4e44 UVfb models. At high emissivity, reionization happens faster, therefore reducing the delay between reionization of the outer and inner progenitor. This then translates to a smaller delay between inner and outer MW sub-haloes. The slopes for our baseline models vary from 123 Myr/dex at low emissivity down to 36 Myr/dex at high emissivity, as shown in Tab. \ref{t:res}. 
However, due to the logarithmic evolution of redshift as a function of time, the slope in $\zreion$ remains relatively constant for all baseline models, and is always close to 1/dex.

\paragraph{Impact of UV feedback}
The $\treion$ gradient of the H7e43 UVfb model is slightly steeper than that of the H7e43 NOfb model. However the latter also features a slightly higher overall emissivity due to the lack of UV feedback. Therefore it is very likely that the difference in slopes between these 2 models are just emissivity-driven. This shows that the gradient found is not a consequence of some peculiar topology of reionization resulting from our implementation of radiative feedback. 

\paragraph{Galaxy to galaxy variations}

\begin{figure*}{h}
\begin{tabular}{cc}
{\includegraphics[width=0.47\linewidth,clip]{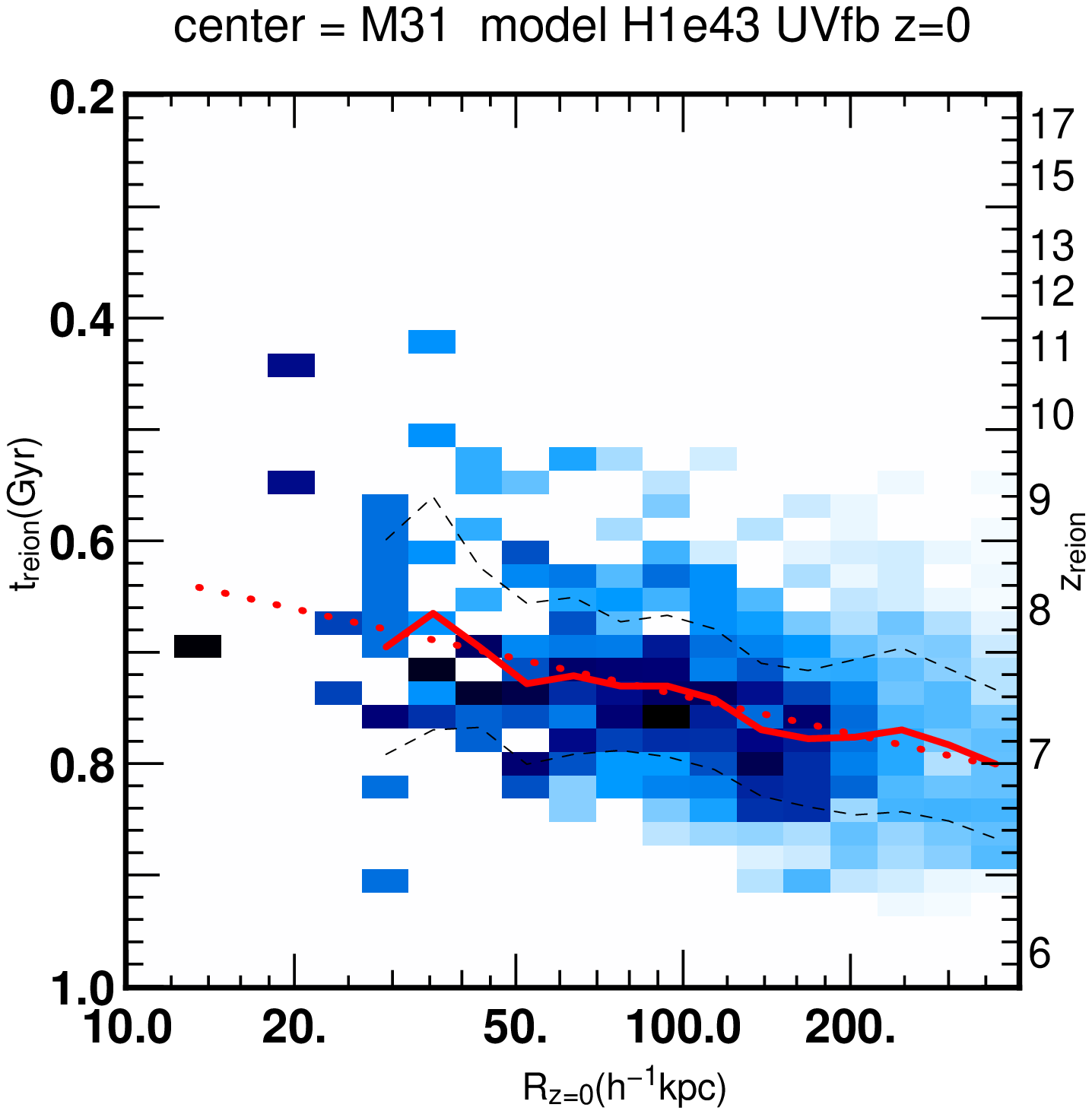}} &
{\includegraphics[width=0.47\linewidth,clip]{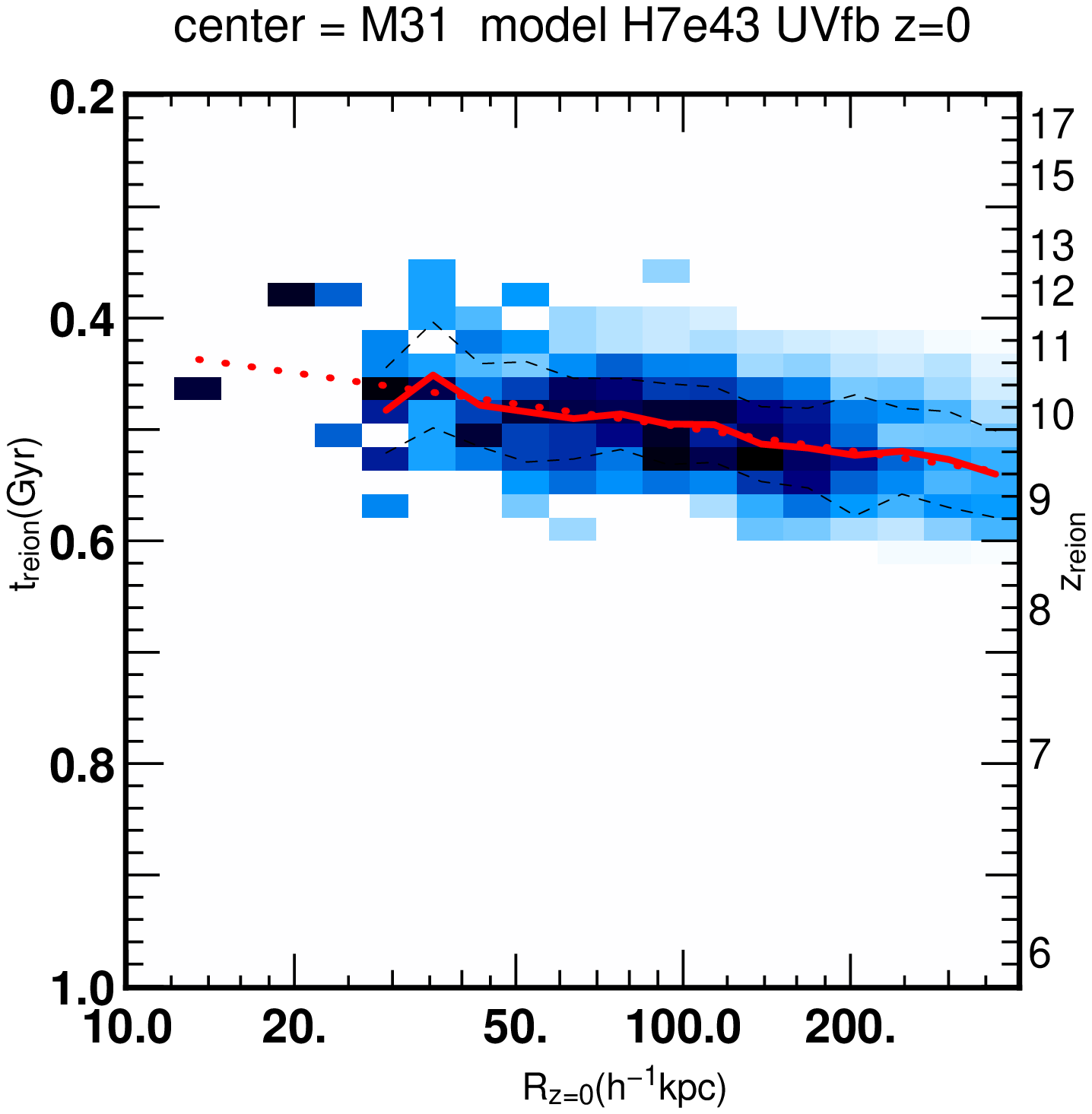}} \\
{\includegraphics[width=0.47\linewidth,clip]{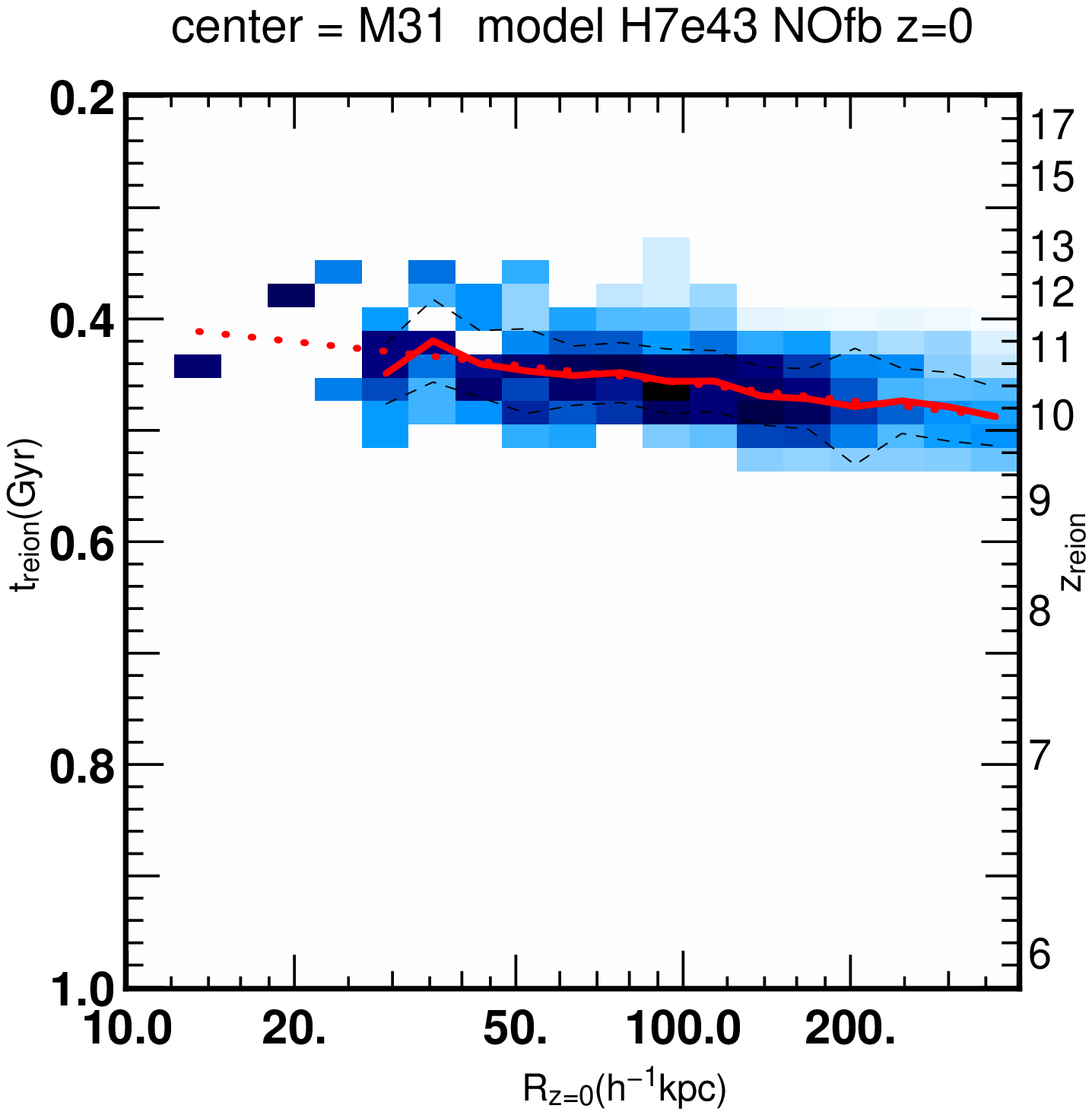}} &
{\includegraphics[width=0.47\linewidth,clip]{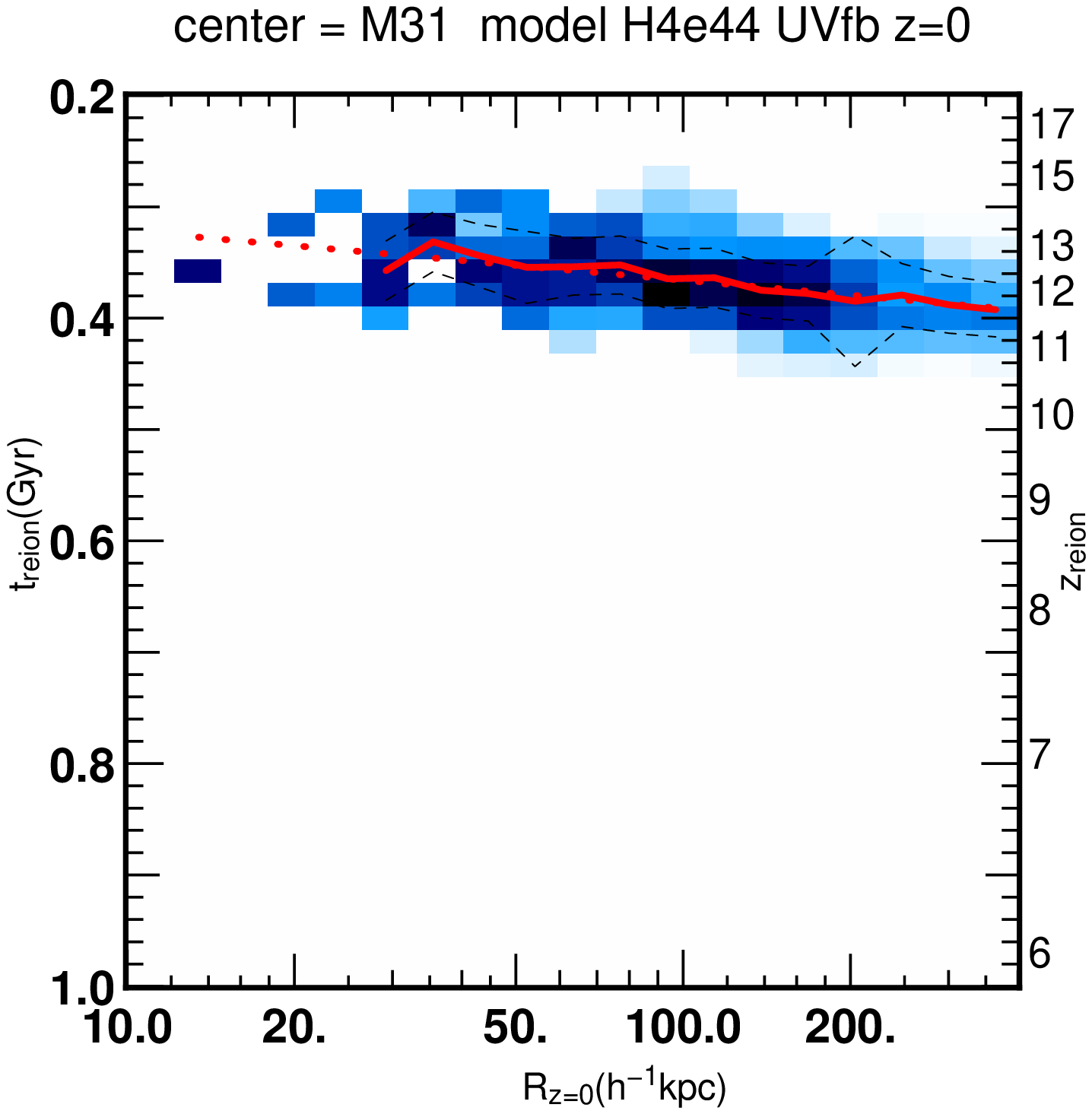}} \\
\end{tabular}
\caption{Same as Fig. \ref{f:rmaplog} for M31.}
\label{f:rmaplogM31}
\end{figure*}

We show the radial lagrangian reionization maps centered on M31 in Fig. \ref{f:rmaplogM31} for our 4 baseline models. They are very similar to the maps centered on the MW, and the same comments as for Fig. \ref{f:rmaplog} hold. The reionization history of the satellites of M31 is correlated with their $z=0$ position as well, with a very similar gradient slope. The average reionization redshift curves (solid line) for MW and M31 are well within $1 \sigma$ of each other. 
It is remarkable that despite presenting strongly different morphologies in eulerian reionization maps, the radial profiles of $\zreion$ at z=0 be so similar for both galaxies. This suggests that the stratification of the satellite population in reionization history is a general feature of all massive galaxies which undergo internal reionization.

\paragraph{Halo versus SPH star model}
\begin{figure*}{t}
\begin{tabular}{cc}
{\includegraphics[width=0.47\linewidth,clip]{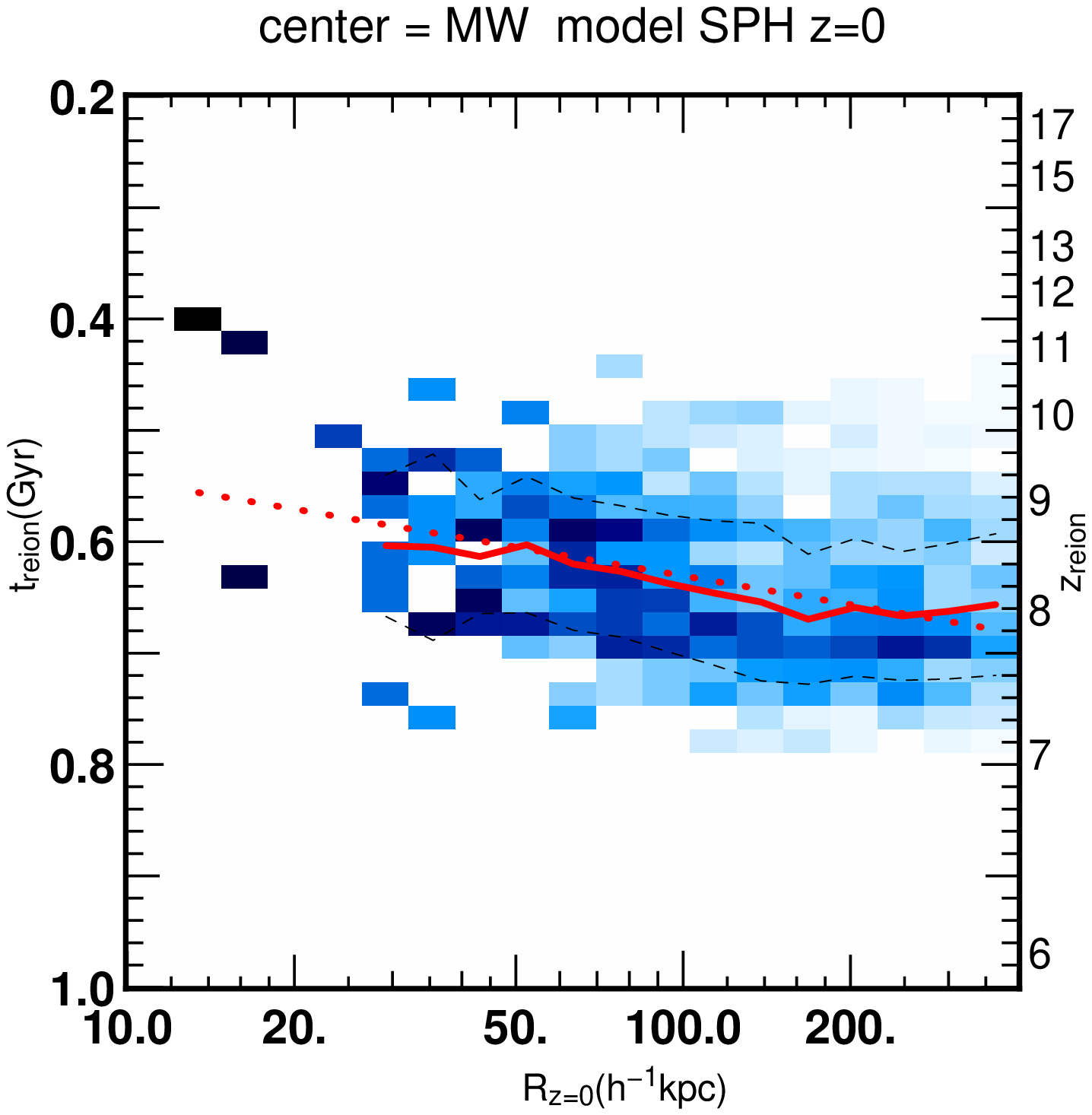}} &
{\includegraphics[width=0.47\linewidth,clip]{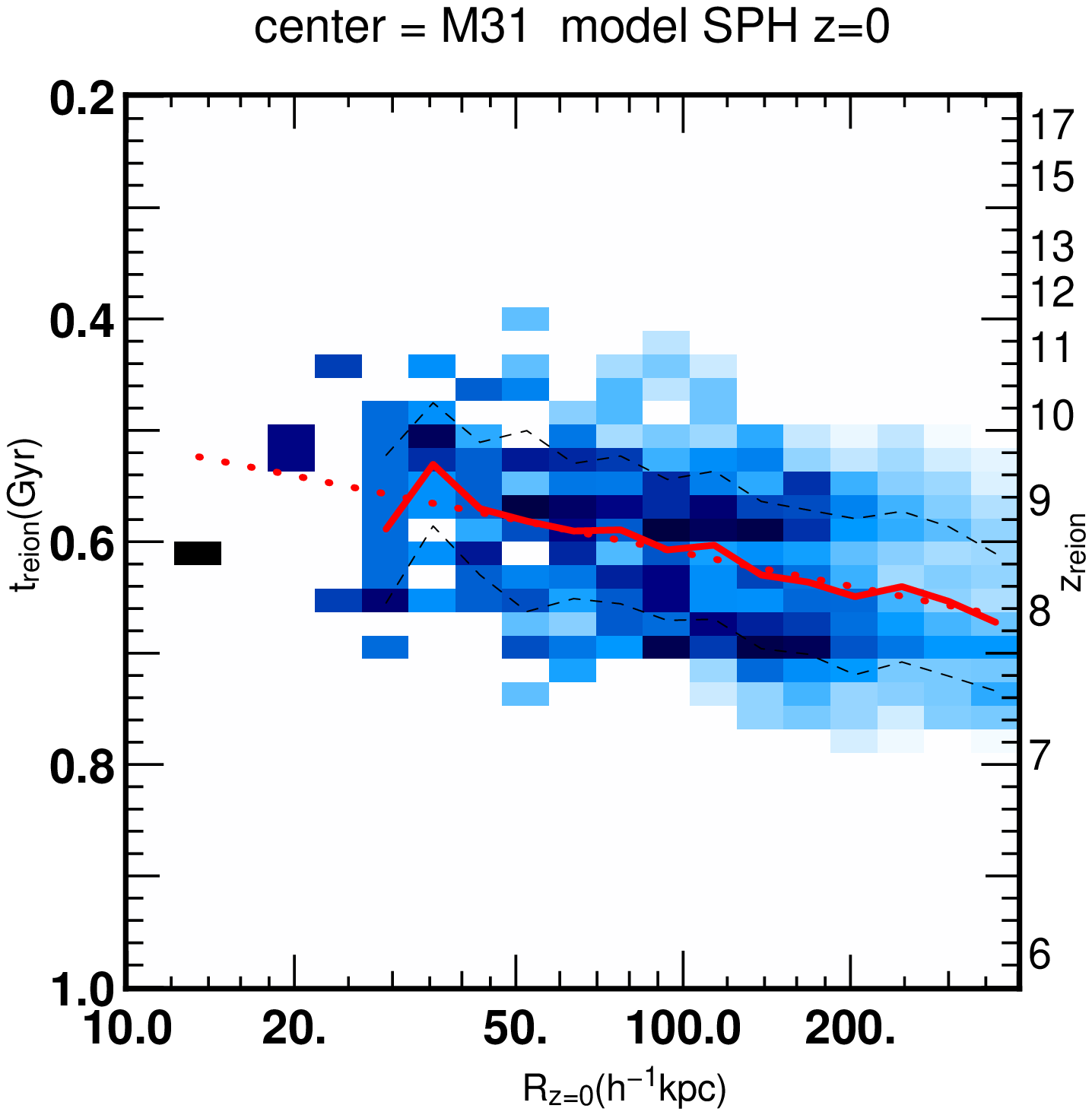}}
\end{tabular}
\caption{Same as Fig. \ref{f:rmaplog} using the SPH model, for MW and M31.}
\label{f:sphgrads}
\end{figure*}

In all of our baseline models, the sources are dark matter haloes, taken from AHF halo catalogs. However, the underlying Gadget-2 simulation from which we take the gas distribution and halo catalogs also features star formation, spawning star particles, which we can thus use as UV sources if young enough. In principle, stars form in haloes, so we expect that at least the location of sources will be the same as with the halo model. However, there may still be a number of significant differences between the two source models:
\begin{itemize}
\item{supernova hydrodynamical feedback is best accounted for in the SPH formalism used, while our baseline models only include radiative feedback of young stars. 
Explicitly accounting for the effect of supernova explosions in our halo model would only reduce the overall star formation efficiency, which is anyway degenerate with respect to other parameters such as specific emissivity and escape fraction. Besides, the H44 SNfb model uses a very strong prescription for supernova feedback, allowing only massive haloes to be sources, while low mass haloes will form no stars at all.}
\item{the star formation rate of haloes in the SPH simulation are sporadic and complex, at variance from our simple halo model where it is simply proportional to halo mass.}
\item{the baseline halo models include radiative feedback, meaning low mass haloes can be shut down by remote, more massive haloes, as far as several 100s of kpc. This will not happen with the SPH star model, where supernova feedback will be purely internal and local.}
\end{itemize}
In order to investigate the impact of these differences, we repeated our experiments, using this time the SPH star particles as sources instead of the halo models. The reionization map for the SPH model was shown in paper I, where we noted very few differences with the halo models of similarly-tuned emissivity, which did not include radiative feedback. Again, we note that the SPH star model reionization map is very similar to that of our baseline models. In terms of timing and structure, it is intermediate between the H1e43 UVfb and H7e43 UVfb models. This also holds for the $\zr$ gradients, as shown in Fig. \ref{f:sphgrads} and Tab. \ref{t:res}. Therefore our main result is robust to a change of source model: there is a clear radial gradient in $\zr$ at z=0 and its slope is correlated with the duration of the reionization process of galaxy progenitor, which is set by the emissivity of the sources.

\subsubsection{Temporal evolution of the gradient}
\begin{figure*}
{\includegraphics[width=0.5\linewidth,clip]{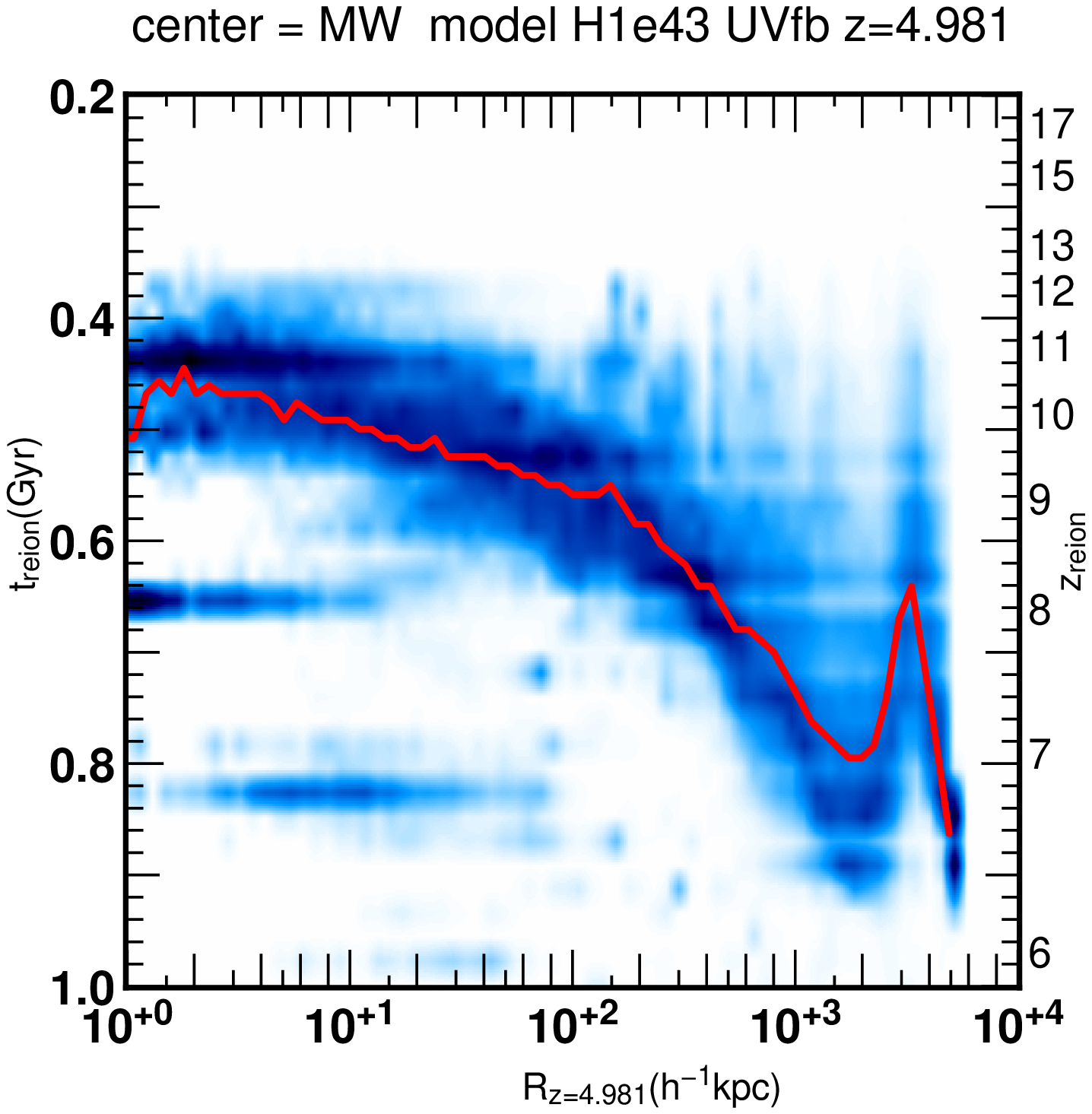}}
{\includegraphics[width=0.5\linewidth,clip]{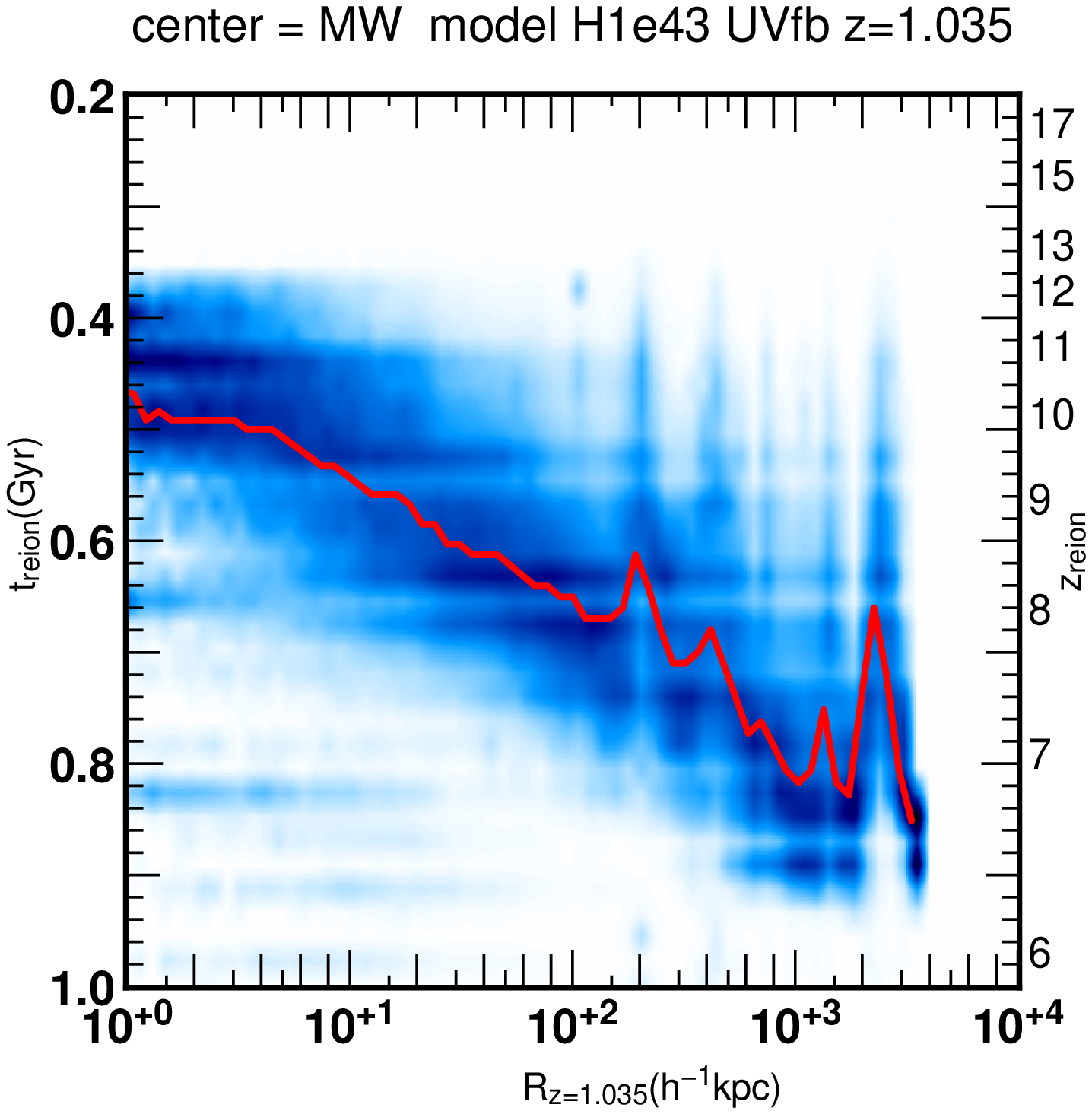}} 
\begin{center}
{\includegraphics[width=0.5\linewidth,clip]{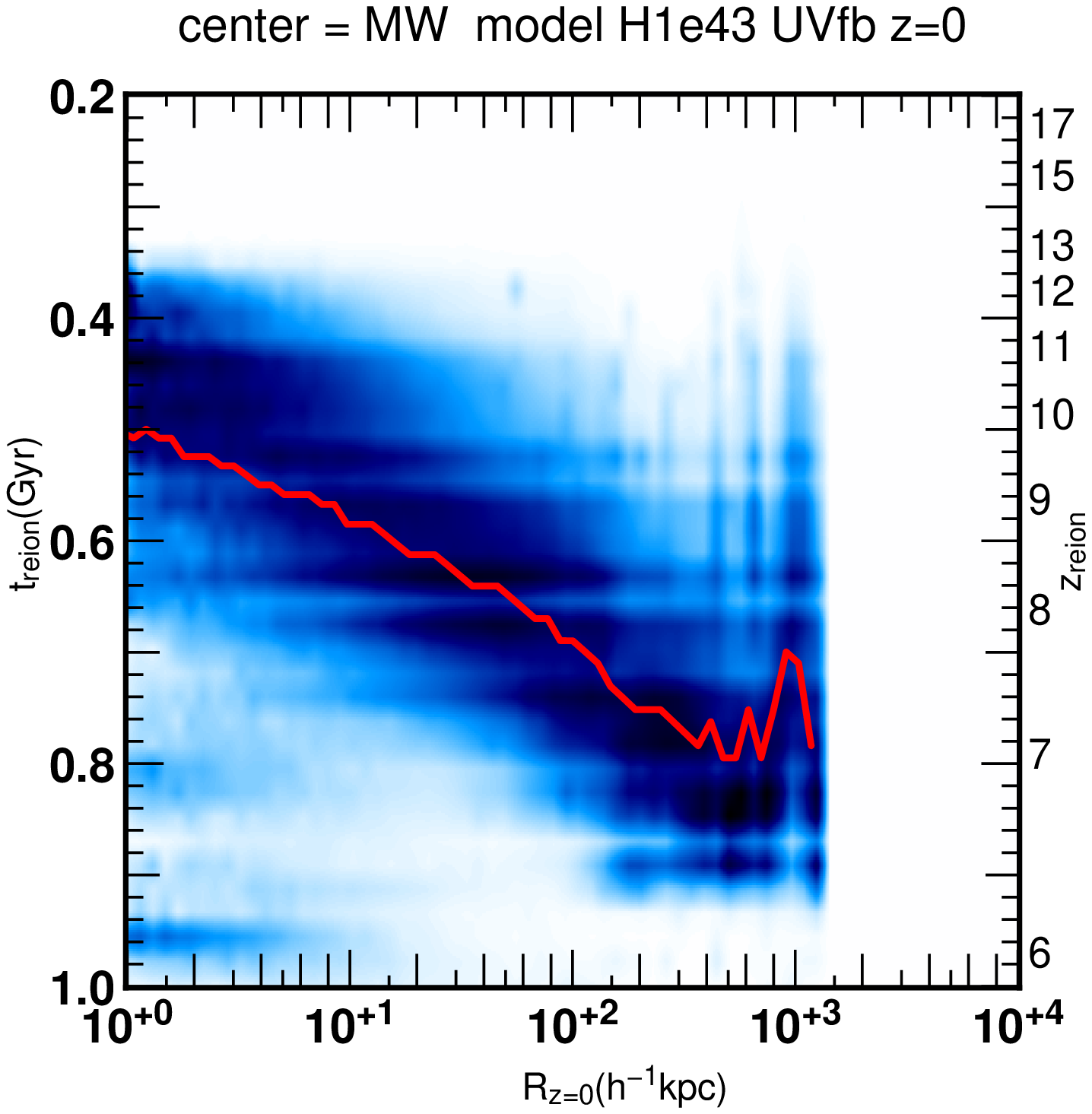}} 
\end{center}
  \caption{Temporal evolution of the reionization redshift gradient of dark matter particles for the H1e43 UVfb model (lowest emissivity), centered on the MW. The title of each panel gives, along with the map center (here MW), the model used and the redshift chosen for the particles positions. Each radial bin is normalized by the number of particles in the bin. The thick red line shows the median $\zr$ of the particles. The center of the MW progenitor (R$=0$) is defined as the center of the main branch halo at all redshifts. The reionization redshift gradient of the particles is steepest at high redshift, and becomes smoother with time. The horizontal ridges are artefacts due to an uneven timing of the RT postprocessing outputs. The distributions were smoothed by a Gaussian for readability.}
\label{f:pipe}
\end{figure*}

In order to gain insight into the origin of the $\zreion$ gradient, we now turn to the investigation of its temporal evolution. Unfortunately, only the most massive sub-haloes at z=0 can be tracked up to $z=6$ and beyond. This severely limits our ability to study the temporal evolution of the gradient. A useful alternative is to simply compute lagrangian reionization maps for all the dark matter particles in the high-resolution region of the simulation. While this is not the same as tracking the halo themselves, we will see that this is still very instructive.
We compute radial lagrangian reionization maps for the H1e43 UVfb model at 3 epochs from just after reionization, z$\sim 5$, to 0, i.e. the only difference between these 3 maps is the positions of the particles at these 3 redshifts.
 The center of the MW progenitor (R$=0$) is defined as the center of the main branch halo at all redshifts \citep{klimentowski2010,srisawat2013}.

The maps are shown in Fig. \ref{f:pipe}. There is a clear flattening of the gradient with time. 
At z=5, very little dynamical evolution has taken place. The gradient is very marked, especially in the 100-1000 \hmkpc range, and the reionization profile is still evocative of a Str\"omgren sphere \citep{barkana2001}, although somewhat perturbed. The center has very few low $\zreion$ particles. 
The dispersion in $\zreion$ is smaller than at any other redshift. At z=1 some mixing within the MW halo and merging of small structures has already taken place, and intermediate $\zreion$ particles have managed to sink in, but the gradient is still rather steep and the dispersion is still smaller than on the z=0 map.
At z=0, the average $\zreion$ profile (thick black line) is much smoother. It has settled in its shallowest slope, which is comparable to the slope found for the satellites on the map of Fig. \ref{f:rmaplog}. The similarity of the $\zr$ - distance distribution between the dark matter particles and the haloes suggests that the process giving rise to the inside-out $\zr$ gradient is the same for both tracers.

Besides the gradient, the maps of Fig. \ref{f:pipe} also show a number of vertical structures (most distinctly at z=1 but also present in the other 2 maps) tracing individual massive haloes. These give rise to the bumps seen in the average $\zr$ profile. While the furthest one is M31 (beyond $1000$ \hmkpc), the other smaller wiggles are produced by smaller objects, although massive enough to reionize internally. The z=1 map features a structure at $\sim 200$ \hmkpc on the verge of merging with the main MW halo. It is not present any more in the z=0 map, signifying the accretion of the object.
We know from paper I and Fig. \ref{f:rmaps} that at least the MW, M31, and to some extent M33 reionize internally, in isolation, with our baseline models, along with a number of more remote, smaller regions. Therefore, the bumps in the average $\zr$ are the counterparts of the internally reionized patches seen in Fig. \ref{f:rmaps}.
In contrast, haloes experiencing external reionization will show up as vertically narrow structures. An example of such an occurence is the region around R$=10^0 - 10^2$ \hmkpc, $\treion \sim 0.825$ Gyr in the z=5 map. This structure has been reionized at z$\sim 6.8$, and accreted on the largest progenitor at some time between z=$6.8 - 5$, as we can see it is already quite extended radially, suggesting a strong tidal interaction. On the later maps, it has migrated inwards and become more and more diffuse (we recall that in these maps, structures can only move horizontally).

As a conclusion, this section confirms our earlier interpretation: inside-out reionization patterns around the brightest sources of the MW progenitor give rise to the $\zreion$ gradients, which are well marked and very steep by the end of reionization.
They are then flattened and blurred by the subsequent 12.7 Gyr of dynamical evolution.
However they are not completely washed out and survive down to present times.



\section{Discussion}
\label{s:discussion}
\subsection{Impact of massive nearby sources}
\label{s:external}
\begin{figure}[t]
{\includegraphics[width=1.1\linewidth,clip]{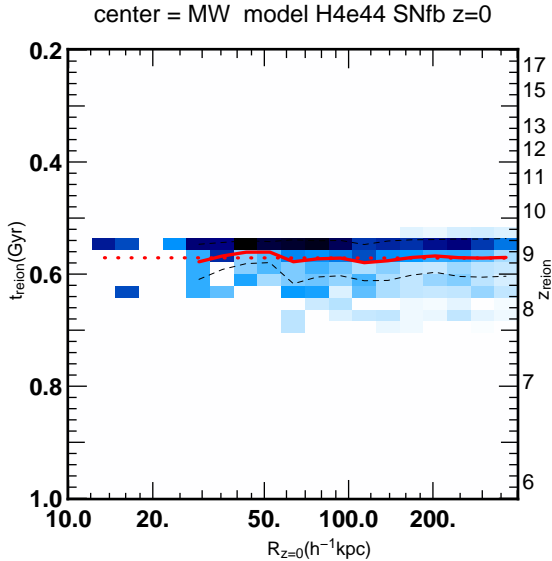}}
\caption{Same as Fig. \ref{f:rmaplog} in case of an external reionization of the MW by M31. The slope of the average $\zreion$  is in this case completely flat.}
\label{f:external}
\end{figure}

An important caveat of our study is that we treat the MW-M31 pair as an isolated system: we do not account for the effect of the nearby galaxy cluster Virgo, which may have been a major source of UV photons during the EoR. Here we show that despite this simplification, our main conclusions hold. Using a larger, lower resolution simulation of the local group formation, I11 showed that for a low emissivity scenario (comparable to our H1e43 and H7e43 models), the MW-M31 system reionizes internally, i.e. its reionization is achieved before the I-front from Virgo reaches us. Therefore, for our H1e43 and H7e43 models, not accounting for Virgo is a reasonable approximation. However, in the high emissivity scenario (corresponding to our H4e44 model), I11 showed that the MW-M31 system is reionized by Virgo. In this case, the reionization of the MW-M31 system is not inside-out, but is driven externally by an I-front from Virgo which sweeps through the MW progenitor halo in less than 15 Myr (estimated from I11 figures), i.e. reionization of the MW is quasi-instantaneous. Therefore no $\zreion$ gradient should be found within the MW halo. Our results show that the high-emissivity scenario leads to an almost flat $\zreion$ gradient (the slope is 3 times smaller than the dispersion in $\zreion$). We see that, based on I11 results, including Virgo could actually make it even flatter.
In order to confirm this expectation, we re-analysed model H44 of paper I (H44 SNfb in this work). The interest of this model for the present study is that it provides us with a case of reionization by an external front: indeed, in this model, paper I showed that the MW progenitor is reionized externally by M31, in a very short time (43 Myr). We computed z=0 the sub-halo $\zreion$ - distance distribution for this model, centered on the MW, shown in Fig. \ref{f:external}, and find the distribution is the flattest of all models.
Therefore, our main conclusion remains unchanged: low emissivity yields a slow, inside-out reionization history for the MW satellite population, whereas high emissivity leads to a fast, quasi-uniform reionization throughout the MW progenitor, resulting in a quasi-constant reionization redshift for the satellite population. 


\subsection{Impact on MW satellites studies}

Our results suggest that the haloes of isolated MW-like galaxies may have a stratified reionization history depending on the emissivity of the sources during the EoR. This is the result of an intrinsical inside-out, fairly slow reionization coupled to some form of weak positional memory of the haloes.
This may have important consequences for the study of galactic satellites: it means that in the low emissivity case, inner satellites at 10 kpc distance (or the disrupted remains thereof) could in average have been reionized up to 180 Myr earlier than the satellites of the outer halo, orbiting at $\sim$ 300 kpc. If indeed reionization is responsible for suppressing star formation in low mass satellites as widely discussed in the literature, then the delayed reionization of the outer halo could lead to differences in stellar content between the inner and outer satellites.
On the contrary, in the high emissivity scenarios, this delay is very small, and therefore the inner and outer satellites would be reionized at the same time. In this case, it would suppress star formation in inner and outer satellites simultaneously, no matter if reionization is internally or externally driven (by Virgo or M31).
Therefore, it seems that by measuring the age of the last generation of stars in a sample of outer and inner satellites, one could discriminate between high/low emissivity scenarios. However, this requires measuring an age difference of the order of 100 Myr in a population older than 10 Gyr, which is extremely challenging even using HST color-magnitude diagrams and stellar population models in the spirit of \cite{dolphin2005}: the width of the oldest age bins in these star formation history reconstructions is typically 10 times larger than the 100 Myr difference we need to measure. 
However, while constrains on the local reionization scenario may be difficult to extract from {\em individual} satellites, the {\em global} properties of the satellite population may still hold important clues. For instance, in the low emissivity case, we find a strong $\zr$ gradient, meaning that the outer satellites may have experienced a longer period of star formation than their more nearby counterparts. Therefore, at a given mass, they could be more luminous and be detected at larger distances. As a consequence, the radial distribution of the satellites could be more extended in the low emissivity case than in the high emissivity case. This recoups the result of \cite{ocvirk2011} which showed using simple semi-analytical modelling that internal reionization, which is more likey at low emissivity than external reionization (I11), produced a more extended radial distribution of the satellites around the MW. The key feature for such a differenciation is the emissivity dependent $\zr$ gradient at z=0, which we confirm with a more realistic modelling in the present paper. 

Therefore it seems that low mass MW satellites such as ultra-faint dwarfs hold important clues about the local reionization history, although understanding and modelling them will be a challenging but exciting task for the years to come. 



\section{Conclusions}
\label{s:conclusions}
We have used high resolution simulations of the formation and reionization of a MW-M31 galaxy pair to investigate the relation between reionization history and present-day position of their satellite population. To do this we have introduced the lagrangian reionization map. It relies on determining a reionization redshift $\zreion$ for each dark matter halo of the simulation. We then explored the relation between the distribution of $\zreion$ and distance to galactic center at z=0, for 4 baseline reionization scenarios featuring various emissivities and feedback processes. In all cases we find that the average $\zreion$ of satellites is higher near galaxy centers (MW and M31). This is due to the inside-out reionization patterns imprinted by massive haloes within the progenitor during the EoR, which end up forming the center of the galaxy. The reionization patterns are slowly flattened by the dynamical evolution of the proto-galaxy and the merging of many substructures. However, they are not totally washed out, and a clear radial gradient in the average satellites reionization redshift still exists today in the halo of our simulated MW and M31, and out to 400 $\hmo$ kpc (571 kpc). 
In the lowest emissivity scenario, the reionization of the outer halo takes place about 180 Myr later than in the inner halo. This is a significant time span compared to the duration of the epoch of reionization itself, and could affect satellite properties by letting remote satellites form stars for longer periods of time or more efficiently than their nearby counterparts. The gradient flattens with increasing source emissivity, because reionization happens faster overall, and becomes spatially quasi-uniform. However, the slope in $\zreion$ remains remarkably constant in all baseline models, at about -1/dex. We checked that our results are robust to changes in the source model by also using the stars spawned by the Gadget-2 simulation as sources rather than the dark matter halos.
In the most luminous scenario, I11 suggests that UV photons from Virgo are expected to speed up drastically the reionization of the MW progenitor. This would likely make any radial $\zreion$ profile even flatter than we predict. We check this by analysing a model in which the MW is externally reionized by M31 and effectively find a flat $\zr$ profile for the satellite population of the MW. On the other hand, the results of the 2 low emissivity scenarios should not be affected by Virgo, since in this regime I11 shows that the local group reionizes internally.
In all cases, the $\zreion$ and $\treion$ gradients are well represented by a linear fit for which we give the parameters. We hope this will help to improve semi-analytical satellite population models \citep{munoz2009,busha2010,maccio2010,li2010,font2011} by allowing authors to implement simply more realistic, position-dependent reionization histories.

As a conclusion, it seems that the population of low mass satellites holds important clues about the local reionization history. However, deciphering these clues is currently very challenging, both from a theoretical and observational standpoint. A huge amount of work remains to be done in order to improve the modelling of these systems, as well as to extend our knowledge of the MW and M31 satellite populations.


\section*{Acknowledgements}
This study was performed in the context of the LIDAU project\footnote{\url{http://aramis.obspm.fr/LIDAU/Site_2/LIDAU_-_Welcome.html}}. The LIDAU project was financed by a French ANR (Agence Nationale de la Recherche) funding (ANR-09-BLAN-0030). The RT computations were performed using HPC resources from GENCI-[CINES/IDRIS] (Grant 2011-[x2011046667]), on the hybrid queue of titane at Centre de Calcul Recherche et Technologie, as well as Curie, during a grand challenge time allocation (project PICON: Photo-Ionisation of CONstrained realizations of the local group). The CLUES simulations were performed at the Leibniz Rechenzentrum Munich (LRZ) and at the Barcelona Supercomputing Center (BSC). 
SG and YH acknowledge support by  DFG grant GO 563/21-1. AK is supported by the {\it Ministerio de Econom\'ia y Competitividad} (MINECO) in Spain through grant AYA2012-31101 as well as the Consolider-Ingenio 2010 Programme of the {\it Spanish Ministerio de Ciencia e Innovaci\'on} (MICINN) under grant MultiDark CSD2009-00064. He also acknowledges support from the {\it Australian Research Council} (ARC) grants DP130100117 and DP140100198. He further thanks Jacques Dutronc for le responsable. The author thanks C. Scannapieco for precious hints dispensed in the initial phase of the project, as well as the CLUES collaborators for useful discussions.  The author thanks D.~Munro for freely distributing his Yorick programming language\footnote{\url{http://www.maumae.net/yorick/doc/index.html}}, and its yorick-gl extension which was used in the course of this study. Finally, we thank the anonymous referee for constructive remarks, which helped improve the paper.

\bibliographystyle{apj}
\bibliography{mybib}
%

\label{lastpage}
\end{document}

%% file: tab4paper2.tex
 \begin{tabular}{cccccccccc} 
 \hline 
 \hline 
 (1) & (2) & (3) & (4) & \multicolumn{2}{c}{(5)} &  \multicolumn{2}{c}{(6)} & \multicolumn{2}{c}{(7)} \\ 
 Model & Source & Rad. & Emissivity & \multicolumn{2}{c}{$\zreion^m$} & \multicolumn{2}{c}{$\Delta {\rm z}_{0.1}^{0.9}$} & \multicolumn{2}{c}{$\Delta {\rm t}$ (Myr)}\\ 
 name & criterion & feedb. & (photons/s/${\rm h^{-1}\Msun}$) & MW & M31 & MW & M31 & MW & M31\\ 
 
 \hline 
 \hline 
 H1e43 UVfb & ${\rm T_{vir}>10^4 K}$ & yes & $ 1.4 \times 10^{43}$ & 8 & 8.1 & 2.9 & 2.81 & 267 & 252 \\ 
 H7e43 UVfb & ${\rm T_{vir}>10^4 K}$ & yes & $ 6.8 \times 10^{43}$ & 10.5 & 10.7 & 2.48 & 2.42 & 126 & 120 \\ 
 H7e43 NOfb & ${\rm T_{vir}>10^4 K}$ & no & $ 6.8 \times 10^{43}$ & 11 & 11.2 & 2.25 & 2.28 & 103 & 101 \\ 
 H7e44 UVfb & ${\rm T_{vir}>10^4 K}$ & yes & $ 6.8 \times 10^{44}$ & 13.1 & 13.3 & 3.27 & 2.14 & 98 & 66 \\ 

 \hline
 SPH & - & no & $ 6.3 \times 10^{45}$$^{\star}$ & 9 & 9.4 & 2.72 & 2.34 & 202 & 159 \\ 

 \hline
 H44 SNfb & ${\rm M>10^9 h^{-1}\Msun}$ & no & $ 4.08 \times 10^{44}$ & 9.1 & 9.7 & 0.55 & 0.32 & 43 & 22 \\ 

 \hline 
 \end{tabular}

%% file: tab4papergrads.tex
 \begin{tabular}{ccccccccc} 
 \hline 
 \hline 
 (1) & (2) & (3) & \multicolumn{2}{c}{(4)} &  \multicolumn{2}{c}{(5)} & \multicolumn{2}{c}{(6)}     \\ 
 Model & Emissivity & Gal.& \multicolumn{2}{c}{Center} & \multicolumn{2}{c}{Gradient} & \multicolumn{2}{c}{Dispersion}  \\ 
 name  & (photons/s/${\rm h^{-1}\Msun}$) & prog. & $\treion^{1kpc}$ &   $\zreion^{1kpc}$ &  ${\mathd {\rm \treion}}/{ \mathd \log {\rm R}}$ &   ${\mathd {\rm \zreion}}/{ \mathd \log {\rm R}}$   &   $\sigma(\treion) $ & $\sigma(\zreion) $ \\ 
 &&&(Myr)&&(Myr/dex)&(/dex)&(Myr)& \\ 
 \hline 
 \hline 
 \multirow{2}{*}{H1e43 UVfb} & \multirow{2}{*}{$ 1.4 \times 10^{43}$} & MW & 484 & 9.84 & 123 & 1.109 & 65 & 0.56 \\ 
  &  & M31 & 515 & 9.47 & 111 & 0.963 & 64 & 0.57 \\ 
 \multirow{2}{*}{H7e43 UVfb} & \multirow{2}{*}{$ 6.8 \times 10^{43}$} & MW & 381 & 12.14 & 59 & 1.034 & 34 & 0.55 \\ 
  &  & M31 & 358 & 12.23 & 69 & 1.076 & 36 & 0.56 \\ 
 \multirow{2}{*}{H7e43 NOfb} & \multirow{2}{*}{$ 6.8 \times 10^{43}$} & MW & 373 & 12.39 & 44 & 0.874 & 26 & 0.49 \\ 
  &  & M31 & 351 & 12.58 & 52 & 0.936 & 29 & 0.51 \\ 
 \multirow{2}{*}{H7e44 UVfb} & \multirow{2}{*}{$ 6.8 \times 10^{44}$} & MW & 294 & 14.59 & 36 & 1.006 & 30 & 0.82 \\ 
  &  & M31 & 277 & 14.81 & 44 & 1.109 & 27 & 0.68 \\ 

 \hline
 \multirow{2}{*}{SPH} & \multirow{2}{*}{$ 6.3 \times 10^{45}$$^{\star}$} & MW & 458 & 10.52 & 85 & 1.005 & 57 & 0.63 \\ 
  &  & M31 & 411 & 10.9 & 99 & 1.094 & 61 & 0.68 \\ 

 \hline
 \multirow{2}{*}{H44 SNfb} & \multirow{2}{*}{$ 4.08 \times 10^{44}$} & MW & 570 & 9.08 & 0 & 0.003 & 27 & 0.3 \\ 
  &  & M31 & 448 & 10.59 & 41 & 0.503 & 23 & 0.27 \\ 
 \hline

  \end{tabular}